\documentclass[aps, prd, reprint, twocolumn, superscriptaddress]{revtex4-1}
\usepackage{amsmath}
\usepackage{amssymb}
\usepackage{graphicx}
\usepackage{color}
\usepackage{times}
\usepackage{euscript}
\usepackage{amsthm}
\usepackage{amsfonts}
\usepackage[english]{babel}
\usepackage{epstopdf}
\usepackage{multirow,makecell,array}
\usepackage{mathtools}
\usepackage{float}
\usepackage[dvipsnames]{xcolor}
\usepackage{graphicx}
\usepackage{revsymb}
\usepackage{siunitx}

\begin{document}

\title{Schwinger pair production in counterpropagating laser pulses: Identifying volume factors}

\author{A.~G.~Tkachev}
\affiliation{Department of Physics, Saint Petersburg State University, Universitetskaya Naberezhnaya 7/9, Saint Petersburg 199034, Russia}
\author{I.~A.~Aleksandrov}
\affiliation{Department of Physics, Saint Petersburg State University, Universitetskaya Naberezhnaya 7/9, Saint Petersburg 199034, Russia}
\affiliation{Ioffe Institute, Politekhnicheskaya Street 26, Saint Petersburg 194021, Russia}
\author{V.~M.~Shabaev}
\affiliation{Department of Physics, Saint Petersburg State University, Universitetskaya Naberezhnaya 7/9, Saint Petersburg 199034, Russia}
\affiliation{National Research Centre ``Kurchatov Institute'' B.P. Konstantinov Petersburg Nuclear Physics Institute, Gatchina, Leningrad district 188300, Russia}

\begin{abstract}
We investigate the nonperturbative process of vacuum pair production in a combination of two counterpropagating linearly polarized laser pulses of a finite spatial extent. By means of the locally-constant field approximation (LCFA), we calculate the total particle yield for the corresponding four-dimensional setup and compare it with the estimates obtained for simplified low-dimensional scenarios. Within the domain where the LCFA is well justified, we examine a combination of two plane-wave pulses, a standing electromagnetic wave, and a spatially uniform oscillating field and demonstrate that at each of these three levels of approximation, one can accurately predict the actual particle number by multiplying the results by properly chosen volume factors depending on the field parameters. We present closed-form expressions for these factors providing universal prescriptions for evaluating the particle yield. Our final formula connecting the spatially uniform setup with the four-dimensional scenario has a relative uncertainty of the level of $10\%$. The explicit correspondences deduced in this study not only prove the relevance of the approximate predictions, but also allow one to quickly estimate the number of pairs for various realistic scenarios without performing multidimensional LCFA integrations.
\end{abstract}

\maketitle

\section{Introduction}\label{sec:intro}

As became clear almost a century ago~\cite{euler_kockel,sauter_1931,heisenberg_euler}, a self-consistent theory of electromagnetic interactions can be formulated only within a many-particle approach permitting elementary processes with a nonconserving number of quanta, i.e., electrons $e^-$, positrons $e^+$, and photons. In the presence of an external background field, these processes can manifest themselves in remarkable nonlinear phenomena which do not occur in classical Maxwell's theory. One of the most staggering effects is the Sauter-Schwinger mechanism of vacuum electron-positron pair production in strong electromagnetic fields~\cite{sauter_1931,heisenberg_euler,schwinger_1951}. The probability of this process is generally suppressed by a small factor $\mathrm{exp} (- \pi E_\text{c}/E_0)$, where $E_0$ is the external electric field strength and $E_\text{c} = m^2c^3/|e\hbar| \sim 10^{16}~\text{V}/\text{cm}$ is the so-called critical field strength ($e$ and $m$ are the electron charge and mass, respectively). This expression indicates that the pair-production mechanism is intrinsically nonperturbative with respect to $E_0$, so by observing this phenomenon, one can probe the effects of quantum electrodynamics (QED) in strong fields in the regime where perturbation theory is no longer applicable.

The Sauter-Schwinger mechanism has not yet been investigated experimentally as its practical observation requires generation of superstrong electromagnetic backgrounds. One of the possible routes to measuring this effect relies on combining several intense laser pulses, which can become feasible in the near future due to a rapid development of the corresponding experimental tools (see recent reviews~\cite{gonoskov_2022,fedotov_review}). Although very simple theoretical estimates can be obtained by means of the Schwinger formula discussed above, it is strongly desirable to accurately evaluate the number of $e^+e^-$ pairs taking into account the spatiotemporal inhomogeneities of the laser setup with the proper preexponential factor. The latter is, in fact, huge since the macroscopic laser fields are focused within the space-time region whose volume is much larger than $(\lambdabar/c)\lambdabar^3$, where $\lambdabar = \hbar/(mc)$ is the reduced Compton wavelength of the electron, which represents one of the natural scales in QED. Over the past decades, we have witnessed a substantial progress in the development of nonperturbative theoretical techniques, which allow one to describe the Sauter-Schwinger effect by means of various numerical approaches (see, e.g., Refs.~\cite{fradkin_gitman_shvartsman,BB_prd_1991,gavrilov_prd_1996,zhuang_1996,ochs_1998,schmidt_1998,kluger_prd_1998,pervushin_skokov,hebenstreit_prd_2010,blaschke_prd_2011,blinne_gies_2014,woellert_prd_2015,blinne_strobel_2016,aleksandrov_prd_2016,li_prd_2017,lv_pra_2018,schneider_prd_2018,huang_2019,huang_prd_2019,aleksandrov_epjst_2020,aleksandrov_kohlfuerst,aleksandrov_kudlis_klochai,esposti_prd_2024}). Nevertheless, all of these methods can basically be employed for addressing quite simple low-dimensional scenarios which imply that the volume prefactor should be partially taken into account via manual multiplication of the numerical results by a proper dimensional number. For instance, a combination of two counterpropagating plane-wave laser pulses is infinite in the transverse plane, so the results should be multiplied by some effective area $S$. Moreover, we note that even within this simplified scenario, computing the number of pairs represents a very challenging task. (see, e.g., Refs.~\cite{lv_pra_2018,ruf_prl_2009,aleksandrov_2017_2,kohlfuerst_prl_2022,kohlfuerst_arxiv_2022_2}). As will be discussed next, for realistic scenarios, one can rely on the local approximations and even perform the calculations in the most undemanding case of a spatially uniform electric field and use then simple closed-form prescriptions to address multidimensional setups.

The aim of the present study is to deduce the necessary volume factors allowing one to map the results of low-dimensional simulations onto the actual $(3+1)$-dimensional setups. In particular, we will address the question of whether one can approximate a combination of two counterpropagating Gaussian laser pulses of a finite duration and spatial size by a uniform time-dependent electric background and then take into account the coordinate dependence by multiplying the results by a certain volume factor $V$ (numerous theoretical studies were based on the corresponding so-called {\it dipole approximation}~\cite{BB_prd_1991,gavrilov_prd_1996,schmidt_1998,kluger_prd_1998,pervushin_skokov,hebenstreit_prd_2010,blaschke_prd_2011,blinne_gies_2014,blinne_strobel_2016,li_prd_2017,huang_2019,huang_prd_2019,aleksandrov_epjst_2020,aleksandrov_kudlis_klochai,aleksandrov_prd_2017_1,olugh_prd_2019,kohlfuerst_prd_2019,hu_arxiv_2024,majczak_arxiv_2024,aleksandrov_kudlis}). Although the momentum distributions of particles are very sensitive to the spatiotemporal structure of the external field, we will demonstrate that the quantitative predictions of the {\it total} number of pairs can indeed be accurately obtained by properly choosing $V$ and will provide simple analytical expressions for this volume factor. On the one hand, our findings will allow one to easily obtain final numerical estimates for the total number of pairs. On the other hand, with the closed-form prescriptions for $V$, one will be able to avoid rather complicated calculations in the case of multidimensional inhomogeneities. Moreover, our final prescriptions can be applied without studying the derivations.

As we are interested in describing the nonperturbative Sauter-Schwinger effect of vacuum pair production, we assume that the external electromagnetic background is sufficiently strong and slowly varying. In this regime, our main theoretical tool will be the locally-constant field approximation (LCFA), where the total particle yield is computed by integrating the constant-field result over time and position space~\cite{bunkin_tugov, narozhny_pla_2004, narozhny_jetpl_2004, bulanov_jetp_2006, dunne_prd_2006, hebenstreit_prdr_2008, bulanov_prl_2010, gavrilov_prd_2017, sevostyanov_prd_2021, aleksandrov_prd_2019_1, aleksandrov_symmetry_2022}. As will be shown below, the domain of the field parameters where the LCFA is well justified is very broad, so it covers many experimentally relevant field configurations.

Successively going from the dipole approximation (DA) to standing-wave and plane-wave approximations (SWA and PWA) and, finally, to the $(3+1)$-dimensional setup, we will derive the volume factors connecting the corresponding scenarios. Combining then our results, we will provide a simple way to estimate the number of pairs produced in the most realistic field configuration by using only the rough DA predictions. It will be demonstrated that this procedure is quite universal to the choice of the laser-pulse profile and ensures the relative uncertainty of less than $10\%$.

The paper has the following structure. In Sec.~\ref{sec:setup} we describe the external field configuration involving two counterpropagating linearly polarized laser pulses and outline the approximate setups which appear within the PWA, SWA, and DA, respectively. In Sec.~\ref{sec:LCFA} we present the LCFA expressions for the total particle yield in each of the four scenarios. Section~\ref{sec:res} contains the main results of our study. Here we derive analytical formulas for the volume factors and assess their accuracy. Finally, we conclude in Sec.~\ref{sec:conclusions}.

Throughout the text, we employ the units $\hbar = c = 1$, $\alpha = e^2/(4\pi) \approx 1/137$.

\section{External field configurations}\label{sec:setup}

A combination of two counterpropagating linearly polarized laser pulses will be described by the following expressions for the electric and magnetic field components:
\begin{eqnarray}
\boldsymbol{E} (t, \boldsymbol{x}) &=& [\mathfrak{E} (\boldsymbol{x}, \omega t - \omega z) + \mathfrak{E} (\boldsymbol{x}, \omega t + \omega z)] \boldsymbol{e}_x,\\ 
\boldsymbol{H} (t, \boldsymbol{x}) &=& [\mathfrak{E} (\boldsymbol{x}, \omega t - \omega z) - \mathfrak{E} (\boldsymbol{x}, \omega t + \omega z)] \boldsymbol{e}_y,
\end{eqnarray}
where $\{ \boldsymbol{e}_i \}$ are the unit vectors along the Cartesian axes, $\boldsymbol{x} = x  \boldsymbol{e}_x + y \boldsymbol{e}_y + z \boldsymbol{e}_z$, $\mathfrak{E} (\boldsymbol{x}, \eta)$ describes an individual laser pulse, and $\omega$ is the corresponding carrier frequency. The most realistic field configuration within this study will involve two {\it Gaussian beams} of the following form:
\begin{eqnarray}
\mathfrak{E}^{(\text{G})}(\boldsymbol{x}, \eta) &=& \frac{E_0}{2} F(\eta) \frac{w_0}{w(z)} \, \mathrm{exp} \bigg [ - \frac{x^2 + y^2}{w^2 (z)} \bigg ] \nonumber \\
{}&\times &\cos \bigg [ \eta + \frac{\omega (x^2 + y^2)}{2 R(z)} - \psi (z) \bigg ],
\label{eq:gauss_profile}
\end{eqnarray}
where 
\begin{eqnarray}
w(z) &=& w_0 \sqrt{1+z^2/z_\text{R}^2}, \\
R(z) &=& z \Big ( 1+ z_\text{R}^2/z^2 \Big ), \\
\psi (z) &=& \mathrm{arctan}~(z/z_\text{R}).
\end{eqnarray}
Here $w_0$ is the waist radius, $z_\text{R} = \pi w_0^2/\lambda$ is the Rayleigh range, and $\lambda = 2\pi /\omega$. The function $F(\eta)$ is a dimensionless smooth envelope function which vanishes for sufficiently large $|\eta|$. Within the paraxial approximation~\eqref{eq:gauss_profile}, we assume $w_0 \gg \lambda$, which is equivalent to the condition $\theta \ll 1$, where $\theta = \lambda/(\pi w_0)$ is the beam divergence. We will specify the corresponding setup $\boldsymbol{E}^{(\text{G})} (t, \boldsymbol{x})$, $\boldsymbol{H}^{(\text{G})} (t, \boldsymbol{x})$ by choosing the values of $E_0$, $\omega$, and $w_0$. The interaction volume is governed by the length scales $\lambda$ and $w_0$.

In order to substantially simplify the structure of the external field, one can neglect the transverse coordinate dependence and set $w_0/w(z) = 1$, $\psi(z) = 0$. In this case, each Gaussian pulse turns into a plane electromagnetic wave which is infinite in the transverse $x$ and $y$ directions but still has a finite size along the $z$ axis~\cite{lv_pra_2018,ruf_prl_2009,aleksandrov_2017_2,kohlfuerst_prl_2022,kohlfuerst_arxiv_2022_2}. Within this {\it plane-wave approximation} (PWA), the resulting external field reads
\begin{eqnarray}
\boldsymbol{E}^{(\text{PWA})} (t,z) &=& \frac{E_0}{2} \Big [ F(\omega t - \omega z) \cos (\omega t - \omega z) \nonumber \\
{} &+& F(\omega t + \omega z) \cos (\omega t + \omega z) \Big ] \boldsymbol{e}_x,\\ 
\boldsymbol{H}^{(\text{PWA})} (t,z) &=& \frac{E_0}{2} \Big [ F(\omega t - \omega z) \cos (\omega t - \omega z) \nonumber \\
{} &-& F(\omega t + \omega z) \cos (\omega t + \omega z) \Big ] \boldsymbol{e}_y. 
\end{eqnarray}

To further simplify the setup, one can completely disregard the spatial dependence of the envelope function, i.e., one can replace $F(\omega t \pm \omega z)$ with $F(\omega t)$. In this case, the external field will represent a standing electromagnetic wave, which is infinite in all of the spatial directions~\cite{lv_pra_2018,woellert_prd_2015,aleksandrov_kohlfuerst,sevostyanov_prd_2021,aleksandrov_prd_2018,peng_prr_2020}. The field configuration within the {\it standing-wave approximation} (SWA) has the following form:
\begin{eqnarray}
\boldsymbol{E}^{(\text{SWA})} (t,z) &=& E_0 F(\omega t) \cos \omega t \cos \omega z \, \boldsymbol{e}_x, \label{eq:E_swa}\\ 
\boldsymbol{H}^{(\text{SWA})} (t,z) &=& E_0 F(\omega t) \sin \omega t \sin \omega z \, \boldsymbol{e}_y. 
\label{eq:H_swa}
\end{eqnarray}

Finally, one can approximate the external field by a spatially homogeneous background assuming that the particles are predominantly produced in the vicinity of the electric-field maxima. Here we set $z = 0$ and obtain
\begin{eqnarray}
\boldsymbol{E}^{(\text{DA})} (t) &=& E_0 F(\omega t) \cos \omega t \, \boldsymbol{e}_x, \label{eq:field_da_E} \\ 
\boldsymbol{H}^{(\text{DA})} (t) &=& 0. 
\end{eqnarray}
This approach will be called the {\it dipole approximation} (DA).

We underline that while the two Gaussian beams (G) are localized along each of the three spatial axes, the PWA setup is finite only with respect to the $z$ axis and the SWA and DA fields are infinite along each of the three directions. This means that while computing the number of $e^+ e^-$ pairs, one has to take into account the corresponding volume factor depending on the approximation chosen. In what follows, we will identify these factors and provide simple prescriptions allowing one to map the results obtained for the four scenarios described above (DA $\to$ SWA $\to$ PWA $\to$ G).

\section{Locally-constant field approximation} \label{sec:LCFA}

Here we will first present a general LCFA expression for the total number of pairs produced and then apply it to the specific field configurations described above.

\subsection{General expression} \label{sec:lcfa_gen}

The LCFA is based on the following closed-form expression for the total number of pairs produced per unit volume and time in the presence of a constant electromagnetic field~\cite{nikishov_constant}:
\begin{equation}
  \frac{dN}{dt d\boldsymbol{x}} \, [\mathcal{E}, \mathcal{H} ] = \frac{e^2 \mathcal{E} \mathcal{H}}{4\pi^2} \coth{\frac{\pi \mathcal{H}}{\mathcal{E}}} \, \mathrm{exp} \bigg ( \! - \frac{\pi E_\text{c}}{\mathcal{E}} \bigg ),
  \label{eq:LCFA_nikishov}
\end{equation}
where
\begin{eqnarray}
  \mathcal{E} &=& \sqrt{\sqrt{\mathcal{F}^2 + \mathcal{G}^2} + \mathcal{F}},\label{eq:E_cal} \\
  \mathcal{H} &=& \sqrt{\sqrt{\mathcal{F}^2 + \mathcal{G}^2} - \mathcal{F}}.\label{eq:H_cal}
\end{eqnarray}
The Lorentz invariant quantities $\mathcal{F}$ and $\mathcal{G}$ are defined via $\mathcal{F} = (\boldsymbol{E}^2 - \boldsymbol{H}^2)/2$ and $\mathcal{G} = \boldsymbol{E}\cdot \boldsymbol{H}$.

Within the LCFA, one plugs the actual spatiotemporal dependence of the external inhomogeneous field into Eq.~\eqref{eq:LCFA_nikishov} and integrates it then over $t$ and $\boldsymbol{x}$ (see Refs.~\cite{bunkin_tugov, narozhny_pla_2004, narozhny_jetpl_2004, bulanov_jetp_2006, dunne_prd_2006, hebenstreit_prdr_2008, bulanov_prl_2010, gavrilov_prd_2017, sevostyanov_prd_2021, aleksandrov_prd_2019_1, aleksandrov_symmetry_2022}):
\begin{equation}
N = \int d^4 \mathrm{x} \, \frac{dN}{dt d\boldsymbol{x}} \, [\mathcal{E} (\mathrm{x}), \mathcal{H} (\mathrm{x})] ,
\label{eq:lcfa_gen}
\end{equation}
where $\mathrm{x} = (t, \, \boldsymbol{x})$.

In the present study, the total number of particles will be evaluated by means of Eq.~\eqref{eq:lcfa_gen}. Let us now discuss the validity of the LCFA. Since the temporal and spatial variations of the external field are governed by the laser frequency $\omega$, the LCFA is generally well justified once $(E_0/E_\text{c})^{3/2} \gg \omega/m$~\cite{sevostyanov_prd_2021}. This criterion was reliably established in the case of a spatially uniform Sauter-like pulse in Ref.~\cite{sevostyanov_prd_2021} (see also Ref.~\cite{aleksandrov_prd_2019_1}) and also confirmed by means of numerical calculations for an oscillating laser field and a two-dimensional setup representing a standing electromagnetic wave~\cite{sevostyanov_prd_2021}. In Appendix~\ref{sec:app_static}, we explicitly derive exactly the same criterion in the case of a static $x$-dependent Sauter-like field. For a realistic laser wavelength of the order of $1~\SI{}{\micro\metre}$, this condition yields $E_0/E_\text{c} \gg 10^{-4}$. As will be shown below the pair production threshold in terms of the ratio $E_0/E_\text{c}$ for this wavelength amounts to several hundredth, so we will be mainly focused on the interval $0.01 \leqslant E_0/E_\text{c} \leqslant 0.1$. Here the LCFA is definitely valid as even $0.01$ is much larger than $10^{-4}$. On the other hand, if we consider the minimal value $E_0 = 0.01 E_\text{c}$ used in our calculations, then the laser frequency and wavelength will have to satisfy $\omega/m \ll 10^{-3}$ and $\lambda \gg 4 \times 10^{-4}~\SI{}{\micro\metre}$, respectively, which is completely realistic. We also note that according to the numerical data presented in Fig.~1 of Ref.~\cite{sevostyanov_prd_2021} in the case of a Sauter pulse, the relative error of the LCFA is less than $3\%$ already for $(m/\omega)(E_0/E_\text{c})^{3/2} = 10$. In our investigation, this parameter has a minimum value of the order of $10^3$ for $E_0 = 0.01 E_\text{c}$ and a laser wavelength of $1~\SI{}{\micro\metre}$. This suggests that the LCFA itself introduces a negligible uncertainty within the whole domain of the field parameters considered in our study. We also mention that the Keldysh parameter $\gamma \equiv m\omega /|eE_0|$ does not govern the accuracy of the LCFA as the condition $\gamma \ll 1$ is weaker than the above criterion~\cite{aleksandrov_prd_2019_1,sevostyanov_prd_2021}.

\subsection{Dipole approximation (DA)} \label{sec:lcfa_da}

If the external field does not depend on the spatial coordinates, then Eq.~\eqref{eq:lcfa_gen} takes the following simple form:
\begin{equation}
 \frac{N^{(\text{DA})}}{V} = \frac{e^2}{4\pi^3} \int \limits_{-\infty}^{+\infty} \! dt \,  E^2(t) \, \mathrm{exp} 
  \left [ - \frac{\pi E_\text{c}}{|E(t)|} \right ],
  \label{eq:LCFA_uniform}
\end{equation}
where $V = \int d\boldsymbol{x}$ is the normalization volume and $E(t)$ is the corresponding electric field strength. In our case,
\begin{equation}
E(t)= E_0 F(\omega t) \cos \omega t.
\end{equation}
Since the field depends on $t$ via the product $\omega t$, one can substitute $\eta = \omega t$ in Eq.~\eqref{eq:LCFA_uniform} and immediately reveal that the right-hand side is inversely proportional to $\omega$. Accordingly, we will discuss the results in terms of the following dimensionless $\omega$-independent quantity:
\begin{equation}
\nu^{(\text{DA})} = \frac{\omega N^{(\text{DA})}}{m^4 V}.
\label{eq:nu_DA}
\end{equation}
Its explicit form is given by
\begin{equation}
\nu^{(\text{DA})} = \frac{1}{4\pi^3} \frac{E_0^2}{E_\text{c}^2} \int \limits_{-\infty}^{+\infty} d\eta f^2 (\eta) \, \mathrm{exp} 
  \left [ - \frac{\pi E_\text{c}}{E_0 |f (\eta) |} \right ],
\label{eq:DA_final}
\end{equation}
where
\begin{equation}
f (\eta) = F(\eta) \cos \eta.
\label{eq:f}
\end{equation}

\subsection{Standing-wave approximation (SWA)} \label{sec:lcfa_swa}

Here we again factor out the $\omega$ dependence:
\begin{equation}
\nu^{(\text{SWA})} = \frac{\omega N^{(\text{SWA})}}{m^4 V}.
\label{eq:nu_swa}
\end{equation}
Let us introduce the following $\omega$-independent function representing a Lorentz invariant:
\begin{equation}
\mathcal{F} (\eta, \xi) = \frac{1}{2} \Big \{ \big [\boldsymbol{E}^{(\text{SWA})} (\eta/\omega, \xi/\omega) \big ]^2 - \big [\boldsymbol{H}^{(\text{SWA})} (\eta/\omega, \xi/\omega) \big ]^2 \Big \}.
\end{equation}
Taking into account Eqs.~\eqref{eq:E_swa} and \eqref{eq:H_swa}, one obtains
\begin{equation}
\mathcal{F} (\eta, \xi) = \frac{E_0^2}{2} \, F^2 (\eta) \cos (\eta + \xi) \cos (\eta - \xi).
\end{equation}
Then we will calculate
\begin{eqnarray}
\nu^{(\text{SWA})} &=& \frac{1}{4\pi^4} \int \limits_{-\infty}^{+\infty} d\eta \int \limits_{-\pi}^{\pi} d\xi \theta (\mathcal{F} (\eta, \xi)) \nonumber \\
{}&\times & \frac{\mathcal{F} (\eta, \xi)}{E_\text{c}^2} \, \mathrm{exp} 
  \left [ - \frac{\pi E_\text{c}}{\sqrt{2\mathcal{F} (\eta, \xi)}} \right ].
\label{eq:swa_final}
\end{eqnarray}
Here the Heaviside step function $\theta$ indicates that $e^+e^-$ pairs are generated only in the space-time regions where the electric field is stronger than the magnetic component. If the latter is not the case, one can Lorentz transform the external field, so that it becomes purely magnetic and, thus, does not produce pairs.

\subsection{Plane-wave approximation (PWA)} \label{sec:lcfa_pwa}

Since the external field is now finite along the $z$ axis, the total number of particles per unit transverse cross sectional area $S$ is also finite. We will compute the following quantity:
\begin{equation}
\nu^{(\text{PWA})} = \frac{\omega^2 N^{(\text{PWA})}}{m^4 S}.
\label{eq:nu_pwa}
\end{equation}
It also does not depend on $\omega$. Here it is convenient to introduce $\eta_- = \omega (t - z)$ and $\eta_+ = \omega (t + z)$. We find
\begin{eqnarray}
\nu^{(\text{PWA})} &=& \frac{1}{8 \pi^3} \frac{E_0^2}{E_\text{c}^2} \int \limits_{-\infty}^{+\infty} d\eta_- \int \limits_{-\infty}^{+\infty} d\eta_+ \theta (f(\eta_-) f(\eta_+)) \nonumber \\
{}&\times & f(\eta_-) f(\eta_+) \, \mathrm{exp} 
  \left [ - \frac{\pi E_\text{c}}{E_0 \sqrt{f(\eta_-) f(\eta_+)}} \right ],
\label{eq:pwa_final}
\end{eqnarray}
where $f (\eta)$ is defined in Eq.~\eqref{eq:f}.

\subsection{Gaussian beams (G)} \label{sec:lcfa_gauss}

The invariant $\mathcal{F}$ is now given by
\begin{equation}
\mathcal{F} = 2 \mathfrak{E} (\boldsymbol{x}, \omega t - \omega z) \mathfrak{E} (\boldsymbol{x}, \omega t + \omega z).
\end{equation}
Let us introduce
\begin{equation}
g (\boldsymbol{x}, \eta) = F(\eta) \cos \bigg [ \eta + \frac{\omega (x^2 + y^2)}{2 R(z)} - \psi (z) \bigg ].
\end{equation}
Then
\begin{eqnarray}
\mathcal{F} &=& \frac{E_0^2}{2} \frac{w_0^2}{w^2(z)} \, \mathrm{exp} \bigg [ - \frac{2(x^2 + y^2)}{w^2 (z)} \bigg ] \nonumber \\
{} & \times & g (\boldsymbol{x}, \omega t - \omega z) g (\boldsymbol{x}, \omega t + \omega z).
\end{eqnarray}
In order to calculate the total number of pairs $N$ according to Eq.~\eqref{eq:lcfa_gen}, we again introduce $\eta_\pm = \omega (t \pm z)$ and also rescale the variables $x$ and $y$ as $x = w_0 \rho_x$, $y = w_0 \rho_y$. Since $z = (\eta_+ - \eta_-)/(2\omega)$, we represent $w(z)$ as
\begin{equation}
w(z) = w_0 \rho (\eta_+ - \eta_-),
\end{equation}
where
\begin{equation}
\rho (\eta) = \sqrt{1 + \frac{\eta^2}{4\omega^2 z_\text{R}^2}} = \sqrt{1 + \bigg ( \frac{\eta}{\omega^2 w_0^2} \bigg )^2}.
\label{eq:rho}
\end{equation}
We now obtain
\begin{eqnarray}
N^{(\text{G})} &=& \frac{1}{8\pi^3} \frac{m^4}{\omega^4} \frac{E_0^2}{E_\text{c}^2} \, (\omega w_0)^2 \int d\rho_x d\rho_y d\eta_- d\eta_+ \nonumber \\
{} & \times & \theta(g (\boldsymbol{x}, \eta_-) g (\boldsymbol{x}, \eta_+)) \frac{1}{\rho^2 (\eta_+ - \eta_-)} \nonumber \\
{} & \times & \mathrm{exp} \bigg [ - \frac{2(\rho_x^2 + \rho_y^2)}{\rho^2 (\eta_+ - \eta_-)} \bigg ] g (\boldsymbol{x}, \eta_-) g (\boldsymbol{x}, \eta_+) \nonumber \\
{} & \times & \mathrm{exp} 
  \Bigg \{ \! - \frac{\pi E_\text{c}}{E_0} \rho (\eta_+ - \eta_-) \, \mathrm{exp} \bigg [ \frac{\rho_x^2 + \rho_y^2}{\rho^2 (\eta_+ - \eta_-)} \bigg ] \nonumber \\
{} & \times & \frac{1}{\sqrt{g (\boldsymbol{x}, \eta_-) g (\boldsymbol{x}, \eta_+)}} \Bigg \},
\end{eqnarray}
where in the functions $g (\boldsymbol{x}, \eta_\pm)$ we imply $x = w_0 \rho_x$, $y = w_0 \rho_y$, and $z = (\eta_+ - \eta_-)/(2\omega)$. Finally, we employ polar coordinates in the $\rho_x \rho_y$ plane, integrate over the polar angle, and substitute $\chi = \rho^2$, where $\rho$ is the corresponding radius, $\rho = \sqrt{\rho_x^2 + \rho_y^2}$. This brings us to the following final expression:
\begin{eqnarray}
N^{(\text{G})} &=& \frac{1}{8\pi^2} \frac{m^4}{\omega^4} \frac{E_0^2}{E_\text{c}^2} \, (\omega w_0)^2 \int \limits _{0}^{+\infty} d\chi \int \limits _{-\infty}^{+\infty} d\eta_- \int \limits _{-\infty}^{+\infty} d\eta_+ \nonumber \\
{} & \times & \theta(\tilde{g} (\chi, z, \eta_-) \tilde{g} (\chi, z, \eta_+)) \frac{1}{\rho^2 (\eta_+ - \eta_-)} \nonumber \\
{} & \times & \mathrm{exp} \bigg [ - \frac{2\chi}{\rho^2 (\eta_+ - \eta_-)} \bigg ] \tilde{g} (\chi, z, \eta_-) \tilde{g} (\chi, z, \eta_+) \nonumber \\
{} & \times & \mathrm{exp} 
  \Bigg \{ \! - \frac{\pi E_\text{c}}{E_0} \rho (\eta_+ - \eta_-) \, \mathrm{exp} \bigg [ \frac{\chi}{\rho^2 (\eta_+ - \eta_-)} \bigg ] \nonumber \\
{} & \times & \frac{1}{\sqrt{\tilde{g} (\chi, z, \eta_-) \tilde{g} (\chi, z, \eta_+)}} \Bigg \},
\label{eq:G_final}
\end{eqnarray}
where
\begin{equation}
\tilde{g} (\chi, z, \eta) \equiv F(\eta) \cos \bigg [ \eta + \frac{\omega w_0^2 \chi}{2 R(z)} - \psi (z) \bigg ]
\label{eq:g_tilde}
\end{equation}
and $z = (\eta_+ - \eta_-)/(2\omega)$.

Here the result is finite and depends on the field parameters $E_0$, $\omega$, and $w_0$. Nevertheless, as can be easily seen, the integrand in Eq.~\eqref{eq:G_final} involves only $E_0$ and the product $\omega w_0  = 2/\theta$, so it is convenient to define
\begin{equation}
\nu^{(\text{G})} = \frac{\omega^4}{m^4} \, N^{(\text{G})},
\label{eq:nu_G}
\end{equation}
which depends only on $E_0$ and $\theta$. As was mentioned above, the paraxial approximation is valid for sufficiently small values of $\theta$, i.e., large $\omega w_0$.

\section{Results} \label{sec:res}

Here we will numerically evaluate Eqs.~\eqref{eq:DA_final}, \eqref{eq:swa_final}, \eqref{eq:pwa_final}, and \eqref{eq:G_final}. By comparing the results of the different approximations, we aim at deriving simple closed-form expressions that establish the corresponding relations among them. In our numerical calculations, we employ the Gaussian envelope
\begin{equation}
F(\eta) = \mathrm{e}^{-(\eta/\sigma)^2},
\label{eq:envelope}
\end{equation}
where $\sigma$ is a dimensionless parameter governing the laser pulse duration.

In what follows, we will separately discuss the transitions DA $\to$ SWA, SWA $\to$ PWA, and PWA $\to$ G.

\subsection{SWA versus DA} \label{sec:da_to_swa}

According to our numerical results, the DA always overestimates the particle yield, which is no surprise as this approximation neglects the spatial profile of the external field by replacing it with unity and, thus, notably increases the local values of the field strength. Within the SWA, the particles are primarily produced in the vicinity of the maxima of the standing-wave electric component, which will be taken into account for identifying the ratio $\nu^{(\text{SWA})}/\nu^{(\text{DA})}$ in the form of simple analytical expressions. To this end, we will proceed in two different directions: we will first develop a simple phenomenological model and then derive the asymptotic expressions for $\nu^{(\text{DA})}$ and $\nu^{(\text{SWA})}$ in the realistic regime $E_0 \ll E_\text{c}$ by means of Laplace's method.

\subsubsection{Phenomenological model} \label{sec:da_to_swa_D}

As was pointed out above, within the SWA, the main contributions to the particle yield arise in the vicinities of the maxima of $\cos \omega z$ in Eq.~\eqref{eq:E_swa}. Let us consider $z \sim 0$ and identify the vicinity $(-\delta, \delta)$ providing the dominant contribution to the particle number within the corresponding cycle of the cosine profile. According to Eq.~\eqref{eq:LCFA_nikishov}, we suggest using the following relation:
\begin{equation}
\cos^2 \omega \delta \, \mathrm{exp} 
  \left ( - \frac{\pi E_\text{c}}{E_0 \cos \omega \delta} \right ) = \zeta \, \mathrm{exp} 
  \left ( - \frac{\pi E_\text{c}}{E_0} \right ),
\label{eq:swa_da_eq}
\end{equation}
which indicates that the particle density $dN/(dtd\boldsymbol{x})$ at $z = \delta$ equals its value at $z=0$ multiplied by a small number $\zeta$. The latter will be determined by fitting our numerical results. Equation~\eqref{eq:swa_da_eq} can be solved numerically yielding the function $\delta = \delta_\zeta (E_0)$. The corresponding ``reduction factor'' is found then via
\begin{equation}
\frac{\nu^{(\text{SWA})}}{\nu^{(\text{DA})}} \approx D_\zeta (E_0) \equiv \frac{2 \delta_\zeta (E_0)}{(2\pi/\omega)} = \frac{\omega \delta_\zeta (E_0)}{\pi}.
\label{eq:D_delta}
\end{equation}
The crucial question here is whether one can find a {\it universal} value of $\zeta$, so that $\nu^{(\text{SWA})}/\nu^{(\text{DA})} \approx D_\zeta (E_0)$ for all $E_0$. By calculating the ratio $\nu^{(\text{SWA})}/\nu^{(\text{DA})}$ numerically, we will identify $\zeta$ which generates the most accurate approximation $D_\zeta (E_0)$. In the weak-field limit $E_0 \ll E_\text{c}$, one can construct an approximate closed-form solution to Eq.~\eqref{eq:swa_da_eq} (see Appendix~\ref{sec:app_WF}):
\begin{multline}
D_\zeta^{(0)} (E_0) = \frac{\sqrt{2|\ln{\zeta}|}}{\pi^{3/2}} \, \sqrt{\frac{E_0}{E_\text{c}}}\\
\times\bigg[1-\frac{1}{\pi}\Big(1-\frac{5}{12}\ln{\zeta}\Big) \, \frac{E_0}{E_\text{c}}+\mathcal{O} \bigg(\frac{E_0^2}{E_c^2}\bigg)\bigg].
\label{eq:D0}
\end{multline}

\begin{figure}[b]
\center{\includegraphics[width=0.99\linewidth]{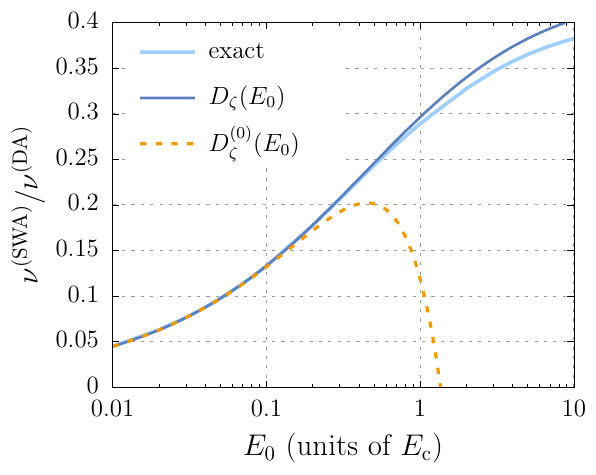}}
\caption{Ratio $\nu^{(\text{SWA})}/\nu^{(\text{DA})}$ evaluated via Eqs.~\eqref{eq:DA_final} and \eqref{eq:swa_final} for $\sigma = 5$ (``exact'') and the functions $D_\zeta (E_0)$ and $D^{(0)}_\zeta (E_0)$ for $\zeta = 0.043$ [see Eqs.~\eqref{eq:D_delta} and \eqref{eq:D0}, respectively].}
\label{fig:swa_da}
\end{figure}

In Fig.~\ref{fig:swa_da} we depict the exact ratio $\nu^{(\text{SWA})}/\nu^{(\text{DA})}$ computed numerically and the functions $D_\zeta (E_0)$ and $D^{(0)}_\zeta (E_0)$ for $\zeta = 0.043$. By exact values we mean those obtained directly by the LCFA expressions~\eqref{eq:DA_final} and \eqref{eq:swa_final}. We observe that the approximate reduction factor $D_\zeta (E_0)$ provides indeed a very accurate prescription for estimating $\nu^{(\text{SWA})}$ by means of $\nu^{(\text{DA})}$. Here we employed $\sigma = 5$, while, according to our results, the ratio displayed in Fig.~\ref{fig:swa_da} is almost independent of the pulse duration (for $\sigma \gtrsim 3$, the curves would be completely indistinguishable in the plot; the curve for $\sigma = 1$ would deviate from that displayed in Fig.~\ref{fig:swa_da} only for $E_0 \gtrsim E_\text{c}$). On the other hand, the quantities $\nu^{(\text{DA})}$ and $\nu^{(\text{SWA})}$ themselves are highly dependent on $\sigma$ at any $E_0$. This suggests that our prescription is indeed quite universal. The leading-order estimate $D^{(0)}_\zeta (E_0)$ [Eq.~\eqref{eq:D0}] can be utilized for $E_0 \lesssim 0.1 E_\text{c}$, so it represents a very useful expression in the realistic domain of subcritical fields. The linear term in the second line of Eq.~\eqref{eq:D0} is rather important here as for $E_0 = 0.1E_\text{c}$ it significantly improves the approximation, so the relative deviation from the exact ratio reduces from $7\%$ to only $1\%$. Our approximations can, of course, be further improved by incorporating the higher-order terms with respect to $E_0/E_\text{c}$, but it appears to be redundant since the accuracy is already very high, so we will opt to relatively simple analytical expressions. We also point out that, e.g., within the interval $0.01 E_\text{c} 
\leqslant E_0 \leqslant 0.1 E_\text{c}$, the quantities $\nu^{(\text{DA})}$ and $\nu^{(\text{SWA})}$ vary over a huge region covering many orders of magnitude (for $E_0 = 0.01 E_\text{c}$ and $\sigma = 5$ we obtain $\nu^{(\text{DA})} \approx 4 \times 10^{-144}$, while for $E_0 = 0.1 E_\text{c}$ we have $\nu^{(\text{DA})} \approx 8\times 10^{-19}$).

The analysis performed in this section also provides additional insights concerning the validity of the LCFA. The condition $(E_0/E_\text{c})^{3/2} \gg \omega/m$ discussed above is equivalent to the requirement that the effective vicinity $\delta$ be much larger than the pair formation length $m/|eE_0|$. This is equivalent to $D_\zeta (E_0) \gg \gamma$, where $\gamma = m\omega/|eE_0|$ is the Keldysh parameter characterizing the laser field. For a realistic laser wavelength of $1~\SI{}{\micro\metre}$ and $E_0 > 0.01 E_\text{c}$, we obtain $\gamma < 3.9 \times 10^{-5}$, which definitely ensures $\gamma \ll D_\zeta (E_0)$ since $D_\zeta (E_0) \gtrsim 0.05$ according to Fig.~\ref{fig:swa_da}.

Although the above prescriptions~\eqref{eq:D_delta} and \eqref{eq:D0} turn out to be very efficient for $\zeta = 0.043$, the origin of the latter value of the fitting parameter is not yet clear as it was merely adjusted by comparing our numerical data. In what follows, we will estimate the ratio $\nu^{(\text{SWA})}/\nu^{(\text{DA})}$ within the weak-field domain $E_0 \ll E_\text{c}$ in a systematic way by means of Laplace's method and show why $\zeta$ should correspond to $\mathrm{e}^{-\pi} \approx 0.0432$. The approach based on Laplace's method will be our main analytical tool throughout the remaining part of the paper.

\subsubsection{Weak-field regime for DA and SWA via Laplace's method}

Since we are primarily interested in the domain of subcritical fields, we can consider the ratio $E_0/E_\text{c}$ as a small parameter and invoke Laplace's method for approximately evaluating the LCFA integrals. Let us begin with the analysis of $\nu^{(\text{DA})}$ in Eq.~\eqref{eq:DA_final}. The main contribution comes from a small vicinity of the origin $\eta = 0$, so we first construct the following expansions:
\begin{eqnarray}
f(\eta) &=& 1 - \bigg ( \frac{1}{2} + \frac{1}{\sigma^2} \bigg ) \eta^2 + \mathcal{O} (\eta^4), \label{eq:series_f} \\
f^2(\eta) &=& 1 - \bigg ( 1 + \frac{2}{\sigma^2} \bigg ) \eta^2 + \mathcal{O} (\eta^4), \label{eq:series_fsqr} \\
\frac{1}{|f(\eta)|} &=& 1 + \bigg ( \frac{1}{2} + \frac{1}{\sigma^2} \bigg ) \eta^2 \nonumber \\
{} &+& \frac{12 + 12\sigma^2 + 5 \sigma^4}{24 \sigma^4} \, \eta^4 + \mathcal{O} (\eta^6). \label{eq:series_1_over_f}
\end{eqnarray}
The last identity immediately leads to
\begin{eqnarray}
\mathrm{exp} \left [ - \frac{\pi E_\text{c}}{E_0 |f (\eta) |} \right ] &=& \mathrm{e}^{- \pi E_\text{c}/E_0} \, \mathrm{exp} \left ( - \frac{\pi E_\text{c}}{E_0} \frac{2+\sigma^2}{2\sigma^2} \, \eta^2 \right ) \nonumber \\
{}&\times& \bigg [ 1 - \frac{\pi E_\text{c}}{E_0} \frac{12 + 12\sigma^2 + 5 \sigma^4}{24 \sigma^4} \, \eta^4 \nonumber \\
{}&+& \mathcal{O} \bigg ( \frac{\eta^6 E_\text{c}}{E_0} \bigg ) + \mathcal{O} \bigg ( \frac{\eta^8 E^2_\text{c}}{E^2_0} \bigg ) \bigg ].
\label{eq:exp_series}
\end{eqnarray}
Here we keep in mind that $\eta \lesssim \sqrt{E_0/E_\text{c}}$ according to the Gaussian profile in the first line of Eq.~\eqref{eq:exp_series}. Now we should only evaluate the integral in Eq.~\eqref{eq:DA_final} using the expansions~\eqref{eq:series_fsqr} and \eqref{eq:exp_series} (the vicinity of $\eta = 0$ can be stretched to the whole real axis). The result reads
\begin{multline}
    \nu^{(\text{DA})} \approx \frac{1}{4\pi^3} \, \sqrt{\frac{2\sigma^2}{2+\sigma^2}} \bigg(\frac{E_0}{E_ \text{c}}\bigg)^{5/2} \mathrm{e}^{- \pi E_\text{c}/E_0} \\
{}\times \bigg [1-\frac{44+44\sigma^2+13\sigma^4}{8\pi(2+\sigma^2)^2} \frac{E_0}{E_\text{c}}+\mathcal{O} \bigg(\frac{E_0^2}{E_\text{c}^2}\bigg)\bigg].
\end{multline}
The $\sigma$-dependent coefficient in the second line rapidly stabilizes with increasing $\sigma$, so we will use a simpler expression,
\begin{multline}
    \nu^{(\text{DA})} \approx \frac{1}{4\pi^3} \, \sqrt{\frac{2\sigma^2}{2+\sigma^2}} \bigg(\frac{E_0}{E_ \text{c}}\bigg)^{5/2} \mathrm{e}^{- \pi E_\text{c}/E_0} \\
{}\times \bigg [1-\frac{13}{8\pi} \frac{E_0}{E_\text{c}}+\mathcal{O} \bigg(\frac{E_0^2}{E_\text{c}^2}\bigg)\bigg].
\label{eq:DA_main}
\end{multline}
As will be demonstrated below, this formula is already very accurate for $E_0 \lesssim 0.1 E_\text{c}$ and sufficiently small $\sigma$ ($\sigma \sim 5$), while for larger values of $\sigma$ it is also necessary to take into account the side maxima of $|f(\eta)|$, i.e., those located near $\eta = \pm \pi$.

First, we easily find that any maximum $\eta_0$ of $|f(\eta)|$ obeys
\begin{equation}
\tan{\eta_0}=-\frac{2\eta_0}{\sigma^2}.
  \label{eq:eta0_eq}
\end{equation}
Since the side maxima play a non-negligible role only for large $\sigma$, we will use the following expansion for the first positive solution of Eq.~\eqref{eq:eta0_eq}:
\begin{equation}
\eta_0 = \pi - \frac{2\pi}{\sigma^2} + \mathcal{O} \bigg(\frac{1}{\sigma^4}\bigg).
  \label{eq:eta0_series}
\end{equation}
The rest part of the calculation is very similar to the above derivations, so we will only present the final expression:
\begin{eqnarray}
    \nu^{(\text{DA})} &\approx& \frac{1}{4\pi^3} \, \sqrt{\frac{2\sigma^2}{2+\sigma^2}} \bigg(\frac{E_0}{E_ \text{c}}\bigg)^{5/2} \nonumber \\
{}&\times& \bigg \{ \mathrm{e}^{- \pi E_\text{c}/E_0} \bigg [1-\frac{13}{8\pi} \frac{E_0}{E_\text{c}}+\mathcal{O} \bigg(\frac{E_0^2}{E_\text{c}^2}\bigg)\bigg] \nonumber \\
{}&+& 2 \big(\mathrm{e}^{-\pi^2/\sigma^2} \big )^{5/2} \, \mathrm{exp} \left ( - \frac{\pi E_\text{c}}{E_0} \, \mathrm{e}^{\pi^2/\sigma^2} \right ) \nonumber \\
{}&\times& \bigg [ 1 + \mathcal{O} \bigg(\frac{E_0}{E_\text{c}}\bigg) + \mathcal{O} \bigg(\frac{1}{\sigma^4}\bigg)\bigg] \bigg \}.
\label{eq:da_laplace_final}
\end{eqnarray}
Here we have disregarded the linear terms with respect to the field strength in the prefactor of the side-maxima contributions. We observe that the additional term coincides with the previously found main contribution~\eqref{eq:DA_main} with a rescaled field amplitude, $E_0 \to E_0 \, \mathrm{exp} (-\pi^2/\sigma^2)$ (it is multiplied by a factor of $2$ since there are two side maxima near $\pm \pi$). This simple prescription is only valid for the leading-order contribution displayed in the third and fourth lines of Eq.~\eqref{eq:da_laplace_final}.

\begin{figure}[t]
\center{\includegraphics[width=0.99\linewidth]{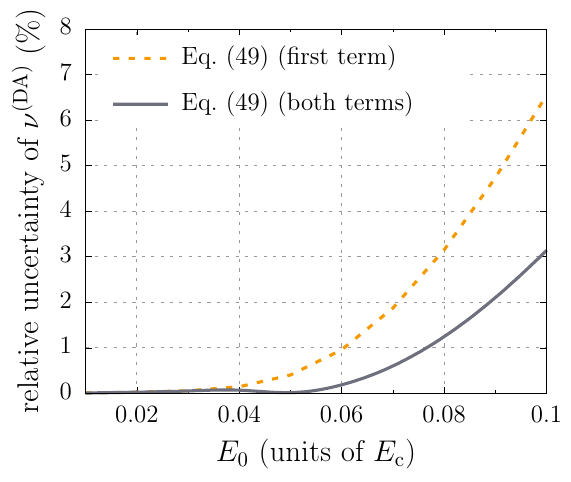}}
\caption{Relative uncertainty of our approximation~\eqref{eq:da_laplace_final} for $\nu^{(\text{DA})}$ as a function of the field amplitude $E_0$ ($\sigma = 10$). The dashed line corresponds to the contribution of the main field maximum, while the solid line incorporates also the side-maxima correction.}
\label{fig:da_laplace_sigma10}
\end{figure}

In Fig.~\ref{fig:da_laplace_sigma10} we present the relative uncertainty of our analytical approximation~\eqref{eq:da_laplace_final} as a function of $E_0$ for $\sigma = 10$. Since it was derived in the weak-field limit, it rapidly breaks down for $E_0 > 0.1 E_\text{c}$, which does not however provide any significant limitation from the viewpoint of the analysis of realistic setups. First, we observe that the accuracy of the approximations is quite high. Second, we clearly see that the side-maxima correction is very important for $\sigma = 10$ as it reduces the relative uncertainty by a factor of $2$, so the latter does not exceed $3\%$. We also point out that in the case $\sigma = 5$ the two analogous curves become indistinguishable and lie below $0.6\%$. For $\sigma = 20$ and $E_0 = 0.1 E_\text{c}$, the main term does not properly capture the effect because the corresponding uncertainty reaches $48\%$. On the other hand, the full expression~\eqref{eq:da_laplace_final} ensures the relative error essentially on the same level as in Fig.~\ref{fig:da_laplace_sigma10} as it does not exceed $5\%$. This clearly indicates that the second term in Eq.~\eqref{eq:da_laplace_final} allows one to keep the errors under control even for large pulse durations~$\sigma$.

\begin{figure}[t]
\center{\includegraphics[width=0.99\linewidth]{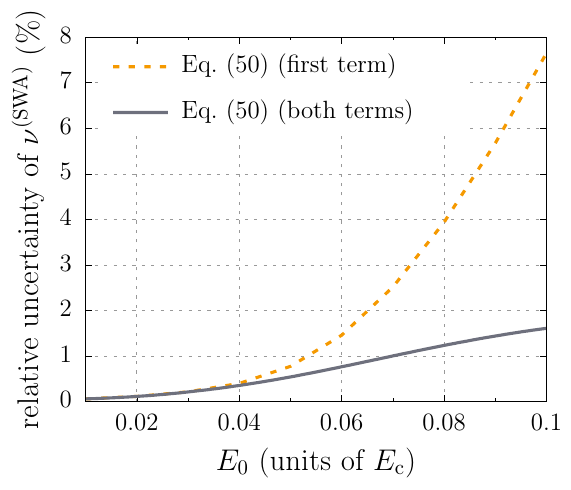}}
\caption{Relative uncertainty of our approximation~\eqref{eq:swa_laplace_final} for $\nu^{(\text{SWA})}$ as a function of $E_0$ ($\sigma = 10$). The dashed line corresponds to the contribution of the main field maximum, while the solid line takes into account also the side-maxima correction.}
\label{fig:swa_laplace_sigma10}
\end{figure}

Let us now turn to the analysis of the SWA. The integrals in Eq.~\eqref{eq:swa_final} can be estimated by means of exactly the same approach as was used above. The final expression incorporating the contributions from the main maximum and the neighboring ones has the following form:
\begin{eqnarray}
    \nu^{(\text{SWA})} &\approx& \frac{1}{2\pi^4} \, \frac{\sigma}{\sqrt{2+\sigma^2}} \bigg(\frac{E_0}{E_ \text{c}}\bigg)^3 \nonumber \\
{}&\times& \bigg \{ \mathrm{e}^{- \pi E_\text{c}/E_0} \bigg [1-\frac{4}{\pi} \frac{E_0}{E_\text{c}}+\mathcal{O} \bigg(\frac{E_0^2}{E_\text{c}^2}\bigg)\bigg] \nonumber \\
{}&+& 2 \big(\mathrm{e}^{-\pi^2/\sigma^2} \big )^{3} \, \mathrm{exp} \left ( - \frac{\pi E_\text{c}}{E_0} \, \mathrm{e}^{\pi^2/\sigma^2} \right ) \nonumber \\
{}&\times& \bigg [ 1 + \mathcal{O} \bigg(\frac{E_0}{E_\text{c}}\bigg) + \mathcal{O} \bigg(\frac{1}{\sigma^4}\bigg)\bigg] \bigg \}.
\label{eq:swa_laplace_final}
\end{eqnarray}
The quality of this approximation is demonstrated in Fig.~\ref{fig:swa_laplace_sigma10}. We reveal basically the same patterns as those already seen in Fig.~\ref{fig:da_laplace_sigma10}, but the side-maxima correction appears here to be even more critical. Moreover, even for $\sigma = 20$, the estimate~\eqref{eq:swa_laplace_final} has the relative error less than $2\%$. In principle, one can straightforwardly take into account the next maxima of the external field, but it is not necessary even for large value of $\sigma = 20$. Next, we will use Eqs.~\eqref{eq:da_laplace_final} and \eqref{eq:swa_laplace_final} to derive a simple analytical approximation for the ratio $\nu^{(\text{SWA})}/\nu^{(\text{DA})}$ and will compare it with Eq.~\eqref{eq:D0}.

\subsubsection{SWA-to-DA ratio via Laplace's method}

Consider the ratio of the expressions~\eqref{eq:da_laplace_final} and \eqref{eq:swa_laplace_final}. The side-maxima corrections to $\nu^{(\text{SWA})}/\nu^{(\text{DA})}$ represent now the difference between the second terms in curly braces in Eqs.~\eqref{eq:swa_laplace_final} and \eqref{eq:da_laplace_final}. The crucial point here is that they differ only in the $\sigma$-dependent exponential factors that satisfy $\mathrm{exp} [5\pi^2/(2\sigma^2)] - \mathrm{exp} (3\pi^2/\sigma^2) = \mathcal{O} (1/\sigma^2)$ for large $\sigma$. This means that the side-maxima corrections to great extent cancel each other out (for small $\sigma$ these corrections are very small compared to the ratio of the main contributions). This explains now why the ratio $\nu^{(\text{SWA})}/\nu^{(\text{DA})}$ was found in Sec.~\ref{sec:da_to_swa_D} to be almost insensitive to varying $\sigma$. Accordingly, our prescription reads
\begin{equation}
\frac{\nu^{(\text{SWA})}}{\nu^{(\text{DA})}}\approx D(E_0) \equiv \frac{\sqrt{2}}{\pi}\sqrt{\frac{E_0}{E_\text{c}}}\bigg(1-\frac{19}{8\pi}\frac{E_0}{E_\text{c}} \bigg).
\label{eq:swa_da_D}
\end{equation}
Let us now compare this expression with Eq.~\eqref{eq:D0}. The leading-order terms in these two formulas coincide if $\zeta = \mathrm{e}^{-\pi} \approx 0.0432$, which explains why this parameter had this particular value in Sec.~\ref{sec:da_to_swa_D}. Nevertheless, the next-order contributions prove to be slightly different: $19/(8\pi) \approx 0.735$ in Eq.~\eqref{eq:swa_da_D} versus $1/\pi + 5/12 \approx 0.756$ in Eq.~\eqref{eq:D0}. In what follows, we will rely on Eq.~\eqref{eq:swa_da_D} since it was obtained in a systematic way.

Our direct numerical calculations revealed that the relative error of Eq.~\eqref{eq:swa_da_D} is indeed very small: even for $E_0 = 0.1 E_\text{c}$ it reaches only $1\%$ independently of $\sigma$.

\subsection{PWA versus SWA} \label{sec:swa_to_pwa}

Our goal here is to take into account the finiteness of the laser pulses, which is neglected within the SWA. The total number of pairs $N^{(\text{PWA})}$ obtained within the PWA is proportional to the cross sectional area $S$, while $N^{(\text{SWA})}$ contains the volume $V = LS$. It is necessary to construct the effective length $L$ as a function of the field parameters. It is natural to represent it as $L = (2 \pi /\omega) N_\text{eff}$, where $N_\text{eff}$ is a dimensionless quantity, which can be viewed as an effective number of cycles in the laser pulse. From Eqs.~\eqref{eq:nu_swa} and \eqref{eq:nu_pwa}, it follows that the condition $N^{(\text{PWA})} = N^{(\text{SWA})}$ is equivalent to
\begin{equation}
\frac{\nu^{(\text{PWA})}}{\nu^{(\text{SWA})}} = 2 \pi N_\text{eff}.
\label{eq:Neff}
\end{equation}
In the realistic subcritical domain $E_0 \lesssim 0.1 E_\text{c}$ and quasistatic regime under consideration, the presence of the Gaussian envelope~\eqref{eq:envelope} makes the pair-production process efficiently occur only within {\it one} carrier cycle in the vicinity of the field maximum [$\eta = 0$ in Eq.~\eqref{eq:envelope}]. Since within the PWA the laser pulses do not fully overlap, one may expect that the correct estimate is $N_\text{eff} \approx 1/2$, so
\begin{equation}
\frac{\nu^{(\text{PWA})}}{\nu^{(\text{SWA})}} \approx \pi.
\label{eq:pwa_swa_1}
\end{equation}
According to our calculations, this is indeed a quite accurate relation once $E_0 \ll E_\text{c}$. However, it can be significantly improved by means of Laplace's method along the lines of our analysis carried out in the previous section.

\begin{figure}[b]
\center{\includegraphics[width=0.99\linewidth]{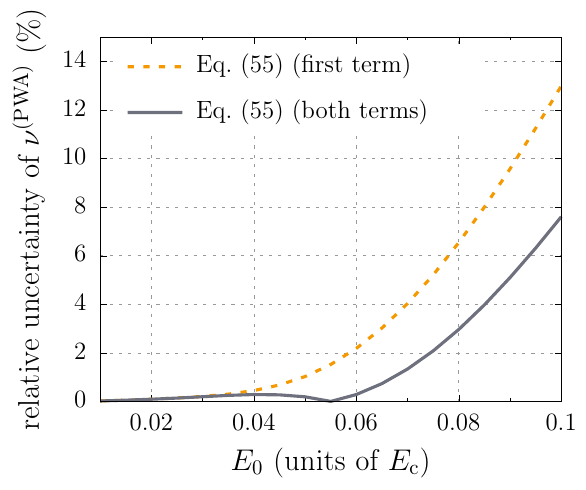}}
\caption{Relative uncertainty of the approximation~\eqref{eq:pwa_laplace_final} for $\nu^{(\text{PWA})}$ as a function of $E_0$ ($\sigma = 10$). The dashed line corresponds to the contribution of the main field maximum, while the solid line incorporates also the side-maxima correction given by the second term in Eq.~\eqref{eq:pwa_laplace_final}.}
\label{fig:pwa_laplace_sigma10}
\end{figure}

Let us first derive an approximate analytical expression for $\nu^{(\text{PWA})}$. First, we consider the main contribution arising from the vicinity of $\eta_+ = \eta_- = 0$ in Eq.~\eqref{eq:pwa_final}. By decomposing the functions $f(\eta_+)$ and $f(\eta_-)$ for small $\eta_\pm$, we obtain the following approximation in the weak-field limit $E_0 \ll E_\text{c}$:
\begin{multline}
    \nu^{(\text{PWA})} \approx \frac{1}{2\pi^3} \, \frac{\sigma^2}{2+\sigma^2} \bigg(\frac{E_0}{E_ \text{c}}\bigg)^3 \mathrm{e}^{- \pi E_\text{c}/E_0} \\
{}\times \bigg [1-\frac{4(3+3\sigma^2+\sigma^4)}{\pi(2+\sigma^2)^2} \frac{E_0}{E_\text{c}}+\mathcal{O} \bigg(\frac{E_0^2}{E_\text{c}^2}\bigg)\bigg] .
\label{eq:pwa_laplace_main}
\end{multline}
Here we will again replace the $\sigma$-dependent fraction in the second line with $4/\pi$ neglecting the terms of the order of $1/\sigma^2$. Taking now into account the side maxima near four points $\eta_\pm = \pm \pi$, we obtain our final prescription,
\begin{eqnarray}
    \nu^{(\text{PWA})} &\approx& \frac{1}{2\pi^3} \, \frac{\sigma^2}{2+\sigma^2} \bigg(\frac{E_0}{E_ \text{c}}\bigg)^3 \nonumber \\
{}&\times& \bigg \{ \mathrm{e}^{- \pi E_\text{c}/E_0} \bigg [1-\frac{4}{\pi} \frac{E_0}{E_\text{c}}+\mathcal{O} \bigg(\frac{E_0^2}{E_\text{c}^2}\bigg)\bigg] \nonumber \\
{}&+& 4 \big(\mathrm{e}^{-\pi^2/\sigma^2} \big )^{3} \, \mathrm{exp} \left ( - \frac{\pi E_\text{c}}{E_0} \, \mathrm{e}^{\pi^2/\sigma^2} \right ) \nonumber \\
{}&\times& \bigg [ 1 + \mathcal{O} \bigg(\frac{E_0}{E_\text{c}}\bigg) + \mathcal{O} \bigg(\frac{1}{\sigma^4}\bigg)\bigg] \bigg \}.
\label{eq:pwa_laplace_final}
\end{eqnarray}
In Fig.~\ref{fig:pwa_laplace_sigma10} we display the relative uncertainty of this formula by comparing its values with the exact numerical results obtained directly by means of Eq.~\eqref{eq:pwa_final}. We confirm again that the contributions from the side maxima should be taken into account for large $\sigma$. This is particularly evident already for $\sigma = 15$ since at $E_0 = 0.1E_\text{c}$, the relative error of the first term in Eq.~\eqref{eq:pwa_laplace_final} reaches no less than $50\%$, whereas the full expression holds the uncertainty of about $8\%$.

To approximate now the ratio $\nu^{(\text{PWA})}/\nu^{(\text{SWA})}$, we divide Eq.~\eqref{eq:pwa_laplace_final} by Eq.~\eqref{eq:swa_laplace_final}. The crucial point here is that the side-maxima contributions no longer cancel each other since they have different numerical prefactors ($4$ and $2$, respectively). It means that the ratio is now rather sensitive to $\sigma$, which can be taken into account by the following expression:
\begin{eqnarray}
    \frac{\nu^{(\text{PWA})}}{\nu^{(\text{SWA})}} &\approx& \frac{\pi \sigma}{\sqrt{2+\sigma^2}} \bigg [ 1 + \frac{2}{\pi \sigma^2} \frac{E_0}{E_\text{c}} \nonumber \\
{}&+& 2 \mathrm{e}^{-3\pi^2/\sigma^2} \, \mathrm{exp} \left ( - \frac{\pi E_\text{c}}{E_0} \frac{\pi^2}{\sigma^2} \right ) \bigg ].
\label{eq:pwa-swa_laplace}
\end{eqnarray}
We observe that for $\sigma \gg 1$ and $E_0 \ll E_\text{c}$, this ratio is indeed close to $\pi$ in accordance with our initial expectation~\eqref{eq:pwa_swa_1}. However, Eq.~\eqref{eq:pwa-swa_laplace} represents a much more accurate approximation as will be now shown via a direct comparison with our numerical LCFA calculations.

\begin{figure}[t]
\center{\includegraphics[width=0.98\linewidth]{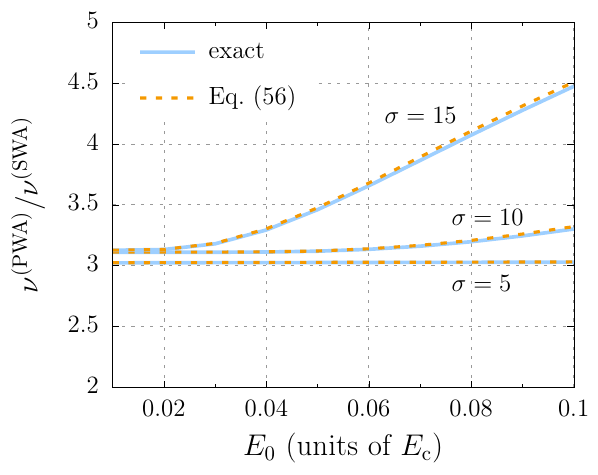}}
\caption{Ratio $\nu^{(\text{PWA})}/\nu^{(\text{SWA})}$ evaluated via Eqs.~\eqref{eq:swa_final} and \eqref{eq:pwa_final} (solid lines) and obtained by means of the approximate prescription~\eqref{eq:pwa-swa_laplace} (dashed lines). In the Gaussian envelope~\eqref{eq:envelope}, we use $\sigma = 5$, $10$, and $15$. The relative uncertainty of these data series is less than $1\%$.}
\label{fig:pwa-swa}
\end{figure}

Our analytical analysis of the pulse shape effects must be justified by direct numerical computations and prove to be universal with respect to the changes in the field parameters. In Fig.~\ref{fig:pwa-swa} we display the exact ratio $\nu^{(\text{PWA})}/\nu^{(\text{SWA})}$ calculated via Eqs.~\eqref{eq:swa_final} and \eqref{eq:pwa_final} and the estimate~\eqref{eq:pwa-swa_laplace} as a function of $E_0$ (note the shift of the origin of the plot). We observe that for large $\sigma$ the ratio $\nu^{(\text{PWA})} / \nu^{(\text{SWA})}$ becomes rather sensitive to $E_0$, which can be taken into account by means of the last term in Eq.~\eqref{eq:pwa-swa_laplace}. Although the relative error in Fig.~\ref{fig:pwa-swa} does not exceed $1\%$, neglecting the side-maxima corrections would lead to large errors that amount, for example, to $6\%$ for $\sigma = 10$ and already to $30\%$ for $\sigma = 15$.

\subsection{Gaussian beams versus PWA} \label{sec:pwa_to_g}

Now our goal is to incorporate the transverse structure of the laser pulses. Let us first provide several preliminary remarks and then invoke Laplace's method as was done in the previous sections. Note that Eqs.~\eqref{eq:nu_pwa} and \eqref{eq:nu_G} lead to the following relation:
\begin{equation}
\frac{\nu^{(\text{G})}}{\nu^{(\text{PWA})}} = \omega^2 S,
\label{eq:nu_G_PWA}
\end{equation}
where we have assumed $N^{(\text{G})} = N^{(\text{PWA})}$. To estimate the effective area $S$, we first point out that it should be proportional to $w_0^2$, which is clear from the analysis of dimensions. This is also in agreement with the fact that $\nu^{(\text{PWA})}$ is $\omega$-independent, while $\nu^{(\text{G})}$ depends only on the product $\omega w_0 = 2/\theta$. Accordingly, the right-hand side of Eq.~\eqref{eq:nu_G_PWA} should read $(\omega w_0)^2 h(E_0/E_\text{c}, \sigma)$, where the function $h$ is to be determined.

The pulse duration $\sigma$ governs the longitudinal structure of the laser field, so we expect that the function $h(E_0/E_\text{c}, \sigma)$ is, in fact, almost insensitive to $\sigma$. On the other hand, the field amplitude $E_0$ strongly affects the volume factor. Its qualitative behavior in the limit $E_0 \ll E_\text{c}$ can be identified by inspecting the Schwinger exponential $\mathrm{exp} (-\pi E_\text{c}/E_0)$, where $E_0$ is modified by the factor $\mathrm{e}^{-\chi}$ [see Eq.~\eqref{eq:G_final}]. The effective radius of the vicinity of $x=y=0$ ($\chi=0$) will possess the same scaling with respect to $E_0$ as $\delta$ in Eq.~\eqref{eq:D_delta}, i.e., it will be proportional to $\sqrt{E_0/E_\text{c}}$ for $E_0 \ll E_\text{c}$, which follows from Eq.~\eqref{eq:D0}. Therefore, the effective area is linear in $E_0$:
\begin{equation}
\frac{\nu^{(\text{G})}}{\nu^{(\text{PWA})}} \approx A (\omega w_0)^2 \, \frac{E_0}{E_\text{c}}.
\label{eq:nu_G_PWA_limit}
\end{equation}
In what follows, we will identify $A$ and refine this approximation via Laplace's method.

Let us now proceed to a systematic analysis of the integrals in Eq.~\eqref{eq:G_final}. It turns out that the three-dimensional integration can also be accurately approximated in the weak-field domain via Laplace's method (see Appendix~\ref{sec:app_G}). The result reads
\begin{eqnarray}
    \nu^{(\text{G})} &\approx& \frac{1}{2\pi^3} \, (\omega w_0)^2 \frac{\sigma^2}{2+\sigma^2} \bigg(\frac{E_0}{E_ \text{c}}\bigg)^4 \nonumber \\
{}&\times& \bigg \{ \mathrm{e}^{- \pi E_\text{c}/E_0} \bigg [1-\frac{8}{\pi} \frac{E_0}{E_\text{c}}+\mathcal{O} \bigg(\frac{E_0^2}{E_\text{c}^2}\bigg)\bigg] \nonumber \\
{}&+& 4 \big(\mathrm{e}^{-\pi^2/\sigma^2} \big )^{4} \, \mathrm{exp} \left ( - \frac{\pi E_\text{c}}{E_0} \, \mathrm{e}^{\pi^2/\sigma^2} \right ) \nonumber \\
{}&\times& \bigg [ 1 + \mathcal{O} \bigg(\frac{E_0}{E_\text{c}}\bigg) + \mathcal{O} \bigg(\frac{1}{\sigma^4}\bigg)\bigg] \bigg \}.
\label{eq:G_laplace}
\end{eqnarray}
Here we highlight the following three features. First, the prefactor contains now the fourth power of $E_0/E_\text{c}$ as we expected in Eq.~\eqref{eq:nu_G_PWA_limit}. Second, the side-maxima correction is similar to that given in Eq.~\eqref{eq:pwa_laplace_final}: it also contains a numerical factor of $4$, but the exponential factor is different. As will be seen below, these corrections significantly shrink each other in the ratio $\nu^{(\text{G})}/\nu^{(\text{PWA})}$. Finally, we observe that the coefficient $8/\pi$ in the second line is twice as large as the analogous factor in Eqs.~\eqref{eq:swa_laplace_final} and \eqref{eq:pwa_laplace_final}. For larger dimensions of the setup, this coefficient also increases, which indicates that the contributions of the higher-order terms in the weak-field expansion may become more important. To assess the accuracy of our prescription~\eqref{eq:G_laplace}, we will now compare it with the exact numerical results.

\begin{figure}[t]
\center{\includegraphics[width=0.99\linewidth]{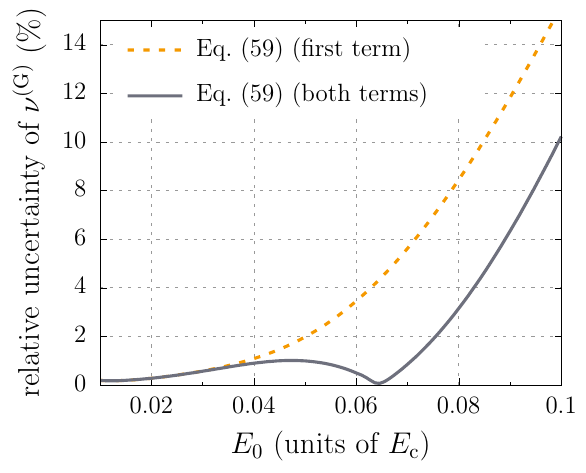}}
\caption{Relative uncertainty of the approximation~\eqref{eq:G_laplace} for $\nu^{(\text{G})}$ as a function of $E_0$ ($\sigma = 10$, $\omega w_0 = 10$). The dashed line corresponds to the contribution of the main field maximum, while the solid line incorporates also the side-maxima correction given by the second term in Eq.~\eqref{eq:G_laplace}.}
\label{fig:G_laplace_sigma10}
\end{figure}

In Fig.~\ref{fig:G_laplace_sigma10} we display $\nu^{(\text{G})}$ as a function of $E_0$ for $\sigma = 10$ and $\omega w_0 = 10$ ($\theta = 0.2$). We note that the parameter $\omega w_0$ does not affect our results, provided $\omega w_0 \gg 1$ as required by the paraxial approximation (we performed the calculations for $5 \leqslant \omega w_0 \leqslant 50$). In Fig.~\ref{fig:G_laplace_sigma10} we reveal essentially the same behavior as was found in Figs.~\ref{fig:swa_laplace_sigma10} and \ref{fig:pwa_laplace_sigma10}. Although we notice a quite nontrivial interplay between the side-maxima correction and the linear term in square brackets in the second line of Eq.~\eqref{eq:G_laplace}, the overall uncertainty does not exceed $10\%$. Nevertheless, the accuracy turns out to be slightly lower for the four-dimensional setup involving Gaussian beams.

\begin{figure}[t]
\center{\includegraphics[width=0.99\linewidth]{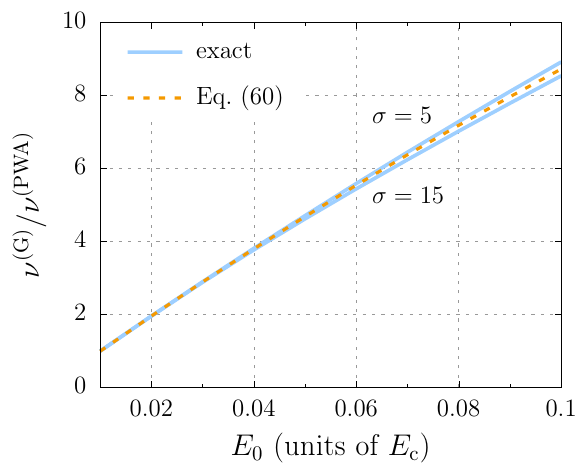}}
\caption{Ratio $\nu^{(\text{G})}/\nu^{(\text{PWA})}$ evaluated via Eqs.~\eqref{eq:pwa_final} and \eqref{eq:G_final} (solid lines) and obtained by means of the approximate expression~\eqref{eq:G_PWA_Laplace} (dashed line) for $\sigma = 5$ and $\sigma = 15$ ($\omega w_0 = 10$). The relative uncertainty proves to be no more than $2\%$.}
\label{fig:g_pwa}
\end{figure}

Let us now discuss the ratio $\nu^{(\text{G})}/\nu^{(\text{PWA})}$, which can be now estimated by means of Eqs.~\eqref{eq:pwa_laplace_final} and \eqref{eq:G_laplace}. As was mentioned above the side-maxima corrections are important only for large $\sigma$, but in the ratio, their difference contains the additional small factor $\mathrm{exp} (4\pi^2/\sigma^2) - \mathrm{exp} (3\pi^2/\sigma^2) = \mathcal{O} (1/\sigma^2)$, so our final approximation will not involve the contributions from the side maxima. It was also pointed out that the linear terms containing $E_0/E_\text{c}$ in Eqs.~\eqref{eq:pwa_laplace_final} and \eqref{eq:G_laplace} combine providing a smaller coefficient $8/\pi - 4/\pi = 4/\pi$. The final expression reads
\begin{equation}
\frac{\nu^{(\text{G})}}{\nu^{(\text{PWA})}} \approx (\omega w_0)^2 \, \frac{E_0}{E_\text{c}} \bigg ( 1 - \frac{4}{\pi} \frac{E_0}{E_\text{c}} \bigg ) .
\label{eq:G_PWA_Laplace}
\end{equation}
This is in agreement with our expectation~\eqref{eq:nu_G_PWA_limit}. In Fig.~\ref{fig:g_pwa} we present the ratio $\nu^{(\text{G})}/\nu^{(\text{PWA})}$ as a function of $E_0$ for $\omega w_0 = 10$ and two different values of $\sigma$. According to our numerical results, this ratio is indeed almost insensitive to $\sigma$ and can be accurately approximated by Eq.~\eqref{eq:G_PWA_Laplace}. The relative error here is always less than $2\%$.

We can now conclude that our approximations of the volume factors, i.e., ratios of the corresponding $\nu$, are always very accurate ensuring the relative uncertainty of less than $2\%$. On the other hand, $\nu$ themselves are predicted not that precisely. This may suggest that combining Eqs.~\eqref{eq:swa_da_D}, \eqref{eq:pwa-swa_laplace}, and \eqref{eq:G_PWA_Laplace}, one will obtain the final DA-to-G formula that should include the sum of the relative errors and, thus, leads to larger uncertainties than those regarding the individual ratios. We will now provide a quantitative analysis of this issue by assessing the accuracy of the final prescription.

\subsection{From DA to Gaussian beams} \label{sec:da_to_gauss}

By combining Eqs.~\eqref{eq:nu_G}, \eqref{eq:swa_da_D}, \eqref{eq:pwa-swa_laplace}, and \eqref{eq:G_PWA_Laplace}, we obtain
\begin{eqnarray}
\frac{N^{(\text{G})}}{\nu^{(\text{DA})}} &\approx& \frac{m^4 w_0^2}{\omega^2} \sqrt{\frac{2\sigma^2}{2+\sigma^2}} \bigg [ 1 - \frac{51}{8\pi} \frac{E_0}{E_\text{c}} \nonumber \\
{}&+& 2 \mathrm{e}^{-3\pi^2/\sigma^2} \, \mathrm{exp} \left ( - \frac{\pi E_\text{c}}{E_0} \frac{\pi^2}{\sigma^2} \right ) \bigg ].
\label{eq:NG_final}
\end{eqnarray}
This formula allows one to obtain the total number of electron-positron pairs produced in the complex $(3+1)$-dimensional setup involving finite Gaussian laser pulses having in hand only the results obtained for the simplest spatially uniform field configuration~\eqref{eq:field_da_E}. To demonstrate the robustness of our prescription, we will now rewrite Eq.~\eqref{eq:NG_final} in a universal form that allows one to address other types of the laser pulse envelope. To this end, we should trace back the origin of the various components of Eq.~\eqref{eq:NG_final}. First, the square root in the first line contains $\sigma$, which should be determined via the Taylor expansion of the envelope function $F(\eta)$, which was chosen in the form~\eqref{eq:envelope}. In the general case, we identify it via
\begin{equation}
\sigma^2 = \frac{2}{|F''(0)|}.
\label{eq:sigma_id}
\end{equation}
For instance, this allows one to fully recover Eq.~\eqref{eq:DA_main} apart from the correction in the second line of this expression. Here we note that the coefficient $51/(8\pi) \approx 2.03$ is not solely governed by the second derivative of $F(\eta)$ since we also took into account the fourth-order terms in the Schwinger exponential factor [see Eq.~\eqref{eq:exp_series}]. Although this coefficient can be expressed in terms of $F''(0)$ and the fourth derivative of $F(\eta)$, we refrain from using such detailed information regarding the pulse profile since it is challenging to experimentally resolve the envelope with such precision. Instead of this, we will replace $51/(8\pi) \to 2$ and confirm the validity of the final formula by means of direct numerical calculations with other envelope shapes. The origin of the term in the second line of Eq.~\eqref{eq:NG_final} is rather trivial. The first exponential factor should be replaced with $F^3 (\pi)$. The last factor involves $\pi^2/\sigma^2$, where $\sigma$ should be obtained by means of Eq.~\eqref{eq:sigma_id}. Here we recall that $\pi^2/\sigma^2$ comes from $1/F(\pi) - 1$, which can be approximated by the Taylor expansion: for large $\sigma$ this expansion is accurate, whereas for small $\sigma$ the whole correction in the second line of Eq.~\eqref{eq:NG_final} is negligible in the weak-field regime. Our final formula reads
\begin{eqnarray}
\frac{N^{(\text{G})}}{\nu^{(\text{DA})}} &\approx& \frac{m^4 w_0^2}{\omega^2} \sqrt{\frac{2\sigma^2}{2+\sigma^2}} \bigg [ 1 - \frac{2E_0}{E_\text{c}} \nonumber \\
{}&+& 2 F^3 (\pi) \, \mathrm{exp} \left ( - \frac{\pi E_\text{c}}{E_0} \frac{\pi^2}{\sigma^2} \right ) \bigg ].
\label{eq:NG_universal}
\end{eqnarray}

\begin{figure}[t]
\center{\includegraphics[width=0.99\linewidth]{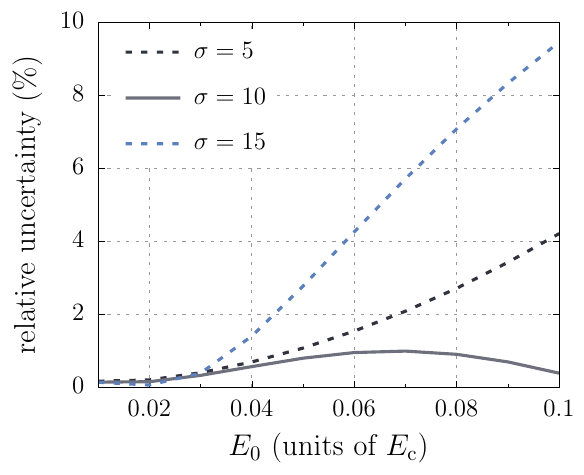}}
\caption{Relative uncertainty of the final prescription~\eqref{eq:NG_universal} as a function of $E_0$ for a Gaussian envelope~\eqref{eq:envelope} and various values of~$\sigma$ ($\omega w_0 = 10$).}
\label{fig:da_g_error}
\end{figure}

The relative error of the final expression~\eqref{eq:NG_universal} is presented in Fig.~\ref{fig:da_g_error} for $\omega w_0 = 10$ and various values of $\sigma$. The uncertainty is less than $10\%$ for all values $E_0 < 0.1 E_\text{c}$. We point out that the error is almost insensitive to $\omega w_0$. The curves displayed in Fig.~\ref{fig:da_g_error} evidently demonstrate a high accuracy of our approximate prescription. Finally, we note that in the case of very short laser pulses $\sigma = 1$, the relative error is always less than $6\%$.

In what follows, we will apply our final prescription~\eqref{eq:NG_universal} to the analysis of two other types of the laser pulse envelope.

\subsection{Universality with respect to the pulse shape} \label{sec:universality}

\begin{figure}[t]
\center{\includegraphics[width=0.99\linewidth]{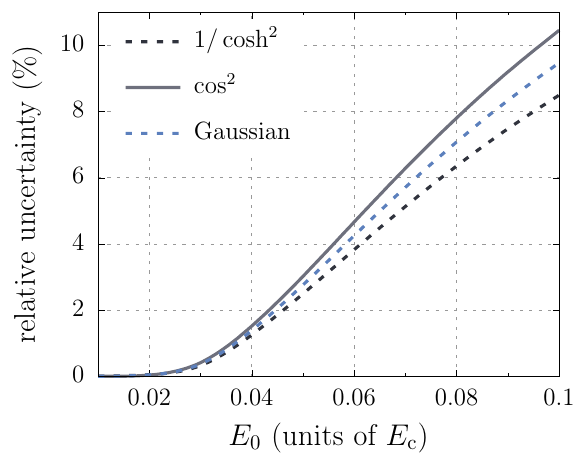}}
\caption{Relative uncertainty of the final approximation~\eqref{eq:NG_universal} as a function of $E_0$ for various types of the pulse envelope [Eqs.~\eqref{eq:envelope}, \eqref{eq:envelope_Sauter}, and \eqref{eq:envelope_cos2}]. The pulse duration is $\sigma = 15$, and $\omega w_0 = 10$.}
\label{fig:env_da_g}
\end{figure}

To demonstrate the efficiency of the final approximation~\eqref{eq:NG_universal} also for other shapes of the pulse envelope, we first perform calculations for a Sauter-type envelope,
\begin{equation}
F(\eta) = \frac{1}{\cosh^2 \eta / \tilde{\sigma}},
\label{eq:envelope_Sauter}
\end{equation}
where $\tilde{\sigma}$ governs the pulse duration. According to Eq.~\eqref{eq:sigma_id}, we identify
\begin{equation}
\sigma = \tilde{\sigma}.
\end{equation}
We now perform the LCFA calculations by means of Eqs.~\eqref{eq:DA_final} and \eqref{eq:G_final} and compare the numerical results with Eq.~\eqref{eq:NG_universal}. The same is also done for the $\cos^2$ envelope:
\begin{equation}
F(\eta) = \cos^2 (\pi \eta/\tilde{\sigma}) \theta (\tilde{\sigma}/2 - |\eta|),
\label{eq:envelope_cos2}
\end{equation}
which is nonzero for $-\tilde{\sigma}/2 < \eta < \tilde{\sigma}/2$. The parameter $\sigma$ defined in Eq.~\eqref{eq:sigma_id} now reads
\begin{equation}
\sigma = \frac{\tilde{\sigma}}{\pi}.
\end{equation}
In Fig.~\ref{fig:env_da_g} we depict the relative errors of Eq.~\eqref{eq:NG_universal} for the three different envelope profiles: Gaussian, Sauter-like, and $\cos^2$. We observe that the numerical uncertainty is almost independent of the pulse shape and always remains on the level of less than $10\%$, which clearly indicates that the final estimates are indeed rather universal. Interestingly, in the case of a Sauter-like $1/\cosh^2$ profile~\eqref{eq:envelope_Sauter}, the accuracy turns out to be even higher than in the case of a Gaussian envelope~\eqref{eq:envelope}.

Finally, let us now provide an explicit numerical example of how Eq.~\eqref{eq:NG_final} can be used to estimate the total particle yield in an experimentally relevant scenario. Suppose we are interested in calculating the total number of pairs for the setup involving two counterpropagating Gaussian pulses with the following parameters: $E_0 = 0.07E_\text{c}$, $\sigma = 5$, $\omega w_0 = 10$ (divergence $\theta = 0.2$). By evaluating Eq.~\eqref{eq:DA_final}, we find $\nu^{(\text{DA})} = 4.44 \times 10^{-25}$. Then by means of Eq.~\eqref{eq:NG_universal}, in terms of $\nu^{(\text{G})} = (\omega/m)^4 N^{(\text{G})}$ we obtain $\nu^{(\text{G})} = 9.62 \times 10^{-24}$. It is very close to the value obtained numerically via Eq.~\eqref{eq:G_final}, which amounts to $\nu^{(\text{G})} = 9.83 \times 10^{-25}$ (the uncertainty is $2.1\%$). To find the total number of electron-positron pairs, one has to multiply $\nu^{(\text{G})}$ by a factor of $(m/\omega)^4$, which governs the 4-volume of the interaction region and represents a huge number since $\omega$ is basically several orders of magnitude smaller than $m$ in realistic scenarios. For instance, if the laser wavelength is $1~\SI{}{\micro\metre}$, then $\omega/m = 3.86 \times 10^{-7}$ and $(m/\omega)^4 = 4.50 \times 10^{25}$. In this case, we obtain $43(1)$ pairs. Our numerical results also yield an accurate estimate for the pair production threshold. It turns out that to observe the Schwinger effect, it is already sufficient to generate two laser pulses with a resulting amplitude of $0.07 E_\text{c}$, which is one order of magnitude smaller than the critical value $E_\text{c}$ and corresponds to the peak intensity of $5.7\times 10^{26}~\text{W}/\text{cm}^2$ for each of the two coherent laser pulses. This value is in agreement with Refs.~\cite{narozhny_jetpl_2004, bulanov_jetp_2006}, where a different model of counterpropagating laser pulses was investigated. We underline that the value $E_0 = 0.07 E_\text{c}$ corresponds to the weak-field regime, where our estimate~\eqref{eq:NG_universal} is well justified. The pair production threshold can be, in principle, lowered due to dynamical effects~\cite{schuetzhold_prl_2008,ilderton_prd_2022,aleksandrov_dase_2022} or by focusing a larger number of laser beams~\cite{bulanov_prl_2010}.

\section{Conclusion} \label{sec:conclusions}

In this study, we deduced simple analytical expressions connecting numerical results for the total particle yield within various approaches to modelling the external field of two counterpropagating laser pulses. Although the existence of an explicit correspondence between the dipole approximation and $(3+1)$-dimensional scenario is nontrivial itself, we derived a universal formula which ensures a relative uncertainty of less than $10\%$. It clearly indicates that the information on the total number of pairs produced by Gaussian beams is already encoded in the simplest setup depending solely on time. It turns out that the envelope function and the oscillating carrier incorporated in the dipole approximation provide one with the necessary estimates, so one should only map them onto the more involved scenario according to the prescription presented in this study. We underline here that the simplified spatially homogeneous setup completely fails to describe the actual momentum distributions of the particles as it disregards the nontrivial dynamics of the created electrons and positrons in the external inhomogeneous background field.

The explicit formulas for the volume factors identified in the present study are expected to be particularly useful for experimentalists as these expressions allow one to easily estimate the particle yield and do not require any specific numerical procedures or analytical calculations [it is not even necessary to carefully follow our derivation of the final formula~\eqref{eq:NG_universal}]. In addition, we demonstrated that the actual pair production threshold in terms of the peak intensity of an individual laser pulse is of the order of $10^{26}~\text{W}/\text{cm}^2$.

\begin{acknowledgments}
We are grateful to the anonymous referees who helped us to significantly improve the manuscript. The study was funded by the Russian Science Foundation, project No.~23-72-01068.
\end{acknowledgments}


\appendix

\section{Validity of the LCFA}\label{sec:app_static}

Here we will discuss the LCFA validity in a bit more detail. Let us first recall a short simple derivation of the criterion $(E_0/E_\text{c})^{3/2} \gg \omega/m$ in the case of a spatially homogeneous electric background $E(t)$~\cite{sevostyanov_prd_2021}. In a subcritical regime $E_0 \ll E_\text{c}$, the main contribution to the particle number arises from the vicinity of the field maximum, where the field strength can be expanded according to
\begin{equation}
E(t) = E_0 - \frac{1}{2}E_0 \Omega^2 t^2 + \ldots
 \label{eq:LCFA_cond_simple_expansion}
\end{equation}
Here we assume that the field maximum is located at $t=0$, and $\Omega^2$ comes from the second derivative of $E(t)$. For the setups considered in our study, $\Omega = \omega \sqrt{1+2/\sigma^2} \approx \omega$. The Schwinger exponential function then reads
\begin{equation}
\mathrm{exp} \left[ \! - \frac{\pi E_\text{c}}{E(t)} \right] \approx \mathrm{exp} \left( \! - \frac{\pi E_\text{c}}{E_0} \right) \mathrm{exp} \left( \! - \frac{\pi E_\text{c} \omega^2 t^2}{2E_0} \right).
 \label{eq:LCFA_cond_simple_schwinger}
\end{equation}
This expression suggests that the effective timescale of the field variations within the pair production process is $\tau_\text{eff} \equiv (1/\omega) \sqrt{E_0/E_\text{c}}$. The external field can be treated as locally constant if $\tau_\text{eff}$ is much larger than the characteristic pair formation length $2m/|eE_0|$. This criterion is equivalent to $(E_0/E_\text{c})^{3/2} \gg \omega/m$.

We see that the above discussion is quite universal and should also hold in the case of multidimensional setups since (a) the formation length remains the same and (b) the spatiotemporal variations of the external field are determined by a single parameter $\omega$ (in principle, the temporal and spatial scales may be different, but this is only possible in specific setups that are not considered in our paper).

To provide a new explicit example that also confirms the above criterion, let us examine a static electric field of the following Sauter-like form:
\begin{equation}
E(x) = \frac{E_0}{\cosh^2 \omega x}.
\label{eq:sauter_static}
\end{equation}
As was comprehensively discussed in Ref.~\cite{gavrilov_prd_2016_gen} the Furry-picture quantization procedure in static fields significantly differs from the case of time-dependent backgrounds. The external field~\eqref{eq:sauter_static} can produce pairs only if $\xi \equiv 1/\gamma = |eE_0|/(m\omega) > 1$, i.e., in the presence of a Klein zone. In this case, the electron number density depends on the transverse momentum $p_\perp = \sqrt{p_y^2+p_z^2}$ and generalized energy $p^0$.

First, we note that the maximal number density corresponds to $p_\perp = p^0 = 0$ and reads~\cite{gavrilov_prd_2016_gen}
\begin{equation}
\frac{\sinh^2 [\pi(m/\omega)\sqrt{\xi^2 - 1}]}{\sinh^2 (\pi \xi m/\omega)}.
\end{equation}
By assuming $\xi \gg 1$ and $m/\omega \gg 1$ and comparing it with the LCFA value $\mathrm{exp} \, (-\pi E_\text{c}/E_0)$, we find that the two results coincide once $(m/\omega)/\xi^3 \ll 1$, which is equivalent to $(E_0/E_\text{c})^{3/2} \gg \omega/m$.

Second, we can directly compare the LCFA predictions for the {\it total} number of pairs with the exact result. According to Eq.~\eqref{eq:lcfa_gen}, the LCFA prediction is exactly the same as in the case of a time-dependent Sauter pulse $E(t) = E_0/\cosh^2 (\omega t)$ (in the static case, the number of pairs is calculated with respect to a unit transverse area $S$ and field duration $T$). The exact result is given by~\cite{gavrilov_prd_2016_gen}
\begin{widetext}
\begin{equation}
\frac{N}{ST} = \frac{1}{\pi^2} \int \limits_0^{p^*_\perp} dp_\perp \, p_\perp \int \limits_{0}^{p^0_*} dp^0 \, \frac{\sinh(\pi p^\text{L}/\omega) \sinh (\pi p^\text{R}/\omega)}{|\sinh \{ (\pi/\omega) [eE_0/\omega + (1/2) (p^\text{L} - p^\text{R})]\} \sinh \{ (\pi/\omega) [eE_0/\omega - (1/2) (p^\text{L} - p^\text{R})] \}|},
\label{eq:N_sauter_static}
\end{equation}
\end{widetext}
where
\begin{eqnarray}
p^*_\perp &=& m\sqrt{\xi^2 -1}, \\
p^0_* &=& m(\xi - \sqrt{1+p_\perp^2/m^2}), \\
p^\text{L} &=& \sqrt{(p^0 + eE/\omega)^2 - m^2 - p_\perp^2}, \\
p^\text{R} &=& \sqrt{(p^0 - eE/\omega)^2 - m^2 - p_\perp^2}.
\end{eqnarray}
By numerically evaluating Eq.~\eqref{eq:N_sauter_static} and computing the relative uncertainty of the LCFA prescription, we came to the same conclusion as that drawn concerning Fig.~1 of Ref.~\cite{sevostyanov_prd_2021}: the numerical results based on the LCFA are accurate if $(E_0/E_\text{c})^{3/2} \gg \omega/m$.

We underline that the parameter $(m/\omega) (E_0/E_\text{c})^{3/2}$ in the main text of our paper exceeds unity by, at least, three orders of magnitude, so the LCFA is extremely accurate and the relative errors of our estimates are solely governed by our approximations of the LCFA integrals.

We also point out that for very strong fields $E_0 \gtrsim E_\text{c}$, the Pauli principle plays a vital role, which is completely neglected by the LCFA~\cite{sevostyanov_prd_2021,aleksandrov_symmetry_2022}. However, it is not relevant to the present study as we focus on the subcritical regime.

Finally, we mention that large $\omega$ can be present in setups regarding dynamically assisted Schwinger effect~\cite{schuetzhold_prl_2008}, where one of the field components can have a very small wavelength. Such scenarios are beyond the scope of the present investigation.


\section{Weak-field limit of Eq.~\eqref{eq:swa_da_eq}}\label{sec:app_WF}

Let us first abbreviate $x=\cos \omega \delta$, $y = \ln \zeta$, $\varepsilon = E_0/(\pi E_\text{c})$ and take a logarithm of Eq.~\eqref{eq:swa_da_eq}:
\begin{equation}
2\ln{x}- \frac{1}{\varepsilon x}  = y - \frac{1}{\varepsilon}.
\label{eq:app_WF_x_eq}
\end{equation}
In the limit $\varepsilon \to 0$, the solution $x$ tends to unity, so we will construct the following expansion:
\begin{equation}
x=1-\sum_{k=1}^{\infty}c_k \varepsilon^k .
\label{eq:app_WF_x_series}
\end{equation}
By substituting it into Eq.~\eqref{eq:app_WF_x_eq}, one can straightforwardly find the following leading coefficients:
\begin{eqnarray}
c_1 &=&-y, \\
c_2 &=&y(2 - y), \\
c_3 &=&-y(1 - y)(4 - y).
\end{eqnarray}
To expand now $\omega \delta = \arccos x$, we note that arccosine has a branch point at $x = 1$ and isolate the square root of $\varepsilon$ according to
\begin{equation}
\omega \delta = \sqrt{\varepsilon} \sum_{k=0}^{\infty}d_k \varepsilon^k .
\label{eq:app_WF_dk}
\end{equation}
By expanding now $\cos \omega \delta$ and employing Eq.~\eqref{eq:app_WF_x_series}, we obtain
\begin{eqnarray}
d_0 &=& \sqrt{2c_1} = \sqrt{2|\ln \zeta|}, \\
d_1 &=& \frac{\sqrt{2}}{12} \, c_1^{3/2} + \frac{c_2}{\sqrt{2c_1}} \nonumber \\
{}&=& -\sqrt{2|\ln \zeta|} \bigg (1 - \frac{5}{12} \, \ln \zeta \bigg),
\end{eqnarray}
which yields then Eq.~\eqref{eq:D0}.


\section{Approximation of the integrals for Gaussian beams}\label{sec:app_G}

Here we derive an approximate closed-form expression for $\nu^{(\text{G})}$, which will allow us to identify the transverse volume factor that establishes the relation between $\nu^{(\text{PWA})}$ and $\nu^{(\text{G})}$. First, we note that the Gouy phase $\psi (z)$ and the term involving $R(z)$ in Eq.~\eqref{eq:g_tilde} do not play a significant role within the LCFA calculations. This means that one can approximately assume $\tilde{g} (\chi, z, \eta) \approx f(\eta)$, where $f(\eta)$ is defined in Eq.~\eqref{eq:f}. Let us represent $\nu^{(\text{G})}$ in the following form:
\begin{widetext}
\begin{equation}
\nu^{(\text{G})} \approx \frac{1}{8 \pi^3} \frac{E_0^2}{E_\text{c}^2} \int \limits_{-\infty}^{+\infty} \! d\eta_- \! \int \limits_{-\infty}^{+\infty} \! d\eta_+ \, \theta (f(\eta_-) f(\eta_+)) f(\eta_-) f(\eta_+) \, \mathrm{exp} 
  \left [ - \frac{\pi E_\text{c}}{E_0 \sqrt{f(\eta_-) f(\eta_+)}} \right ] \Xi (\eta_-, \eta_+, E_0/(\pi E_\text{c}), \omega^2 w_0^2),
\label{eq:app_nu_G}
\end{equation}
where
\begin{eqnarray}
\Xi (\eta_-, \eta_+, \varepsilon, \beta) &=& \pi \beta \int \limits_0^{+\infty} d\chi \, \frac{1}{\rho^2 (\eta_+ - \eta_-, \beta)} \, \mathrm{e}^{-2\chi/\rho^2 (\eta_+ - \eta_-, \beta)} \nonumber \\
{}&\times& \mathrm{exp} 
  \left \{ - \frac{1}{\varepsilon \sqrt{f(\eta_-) f(\eta_+)}} \bigg [ \rho (\eta_+ - \eta_-, \beta) \, \mathrm{e}^{\chi/\rho^2 (\eta_+ - \eta_-, \beta)} - 1 \bigg ] \right \}.
\end{eqnarray}
To explicitly indicate the $\beta$ dependence of the function~\eqref{eq:rho}, we have also introduced
\begin{equation}
\rho (\eta, \beta) = \sqrt{1 + \frac{\eta^2}{\beta^2}}.
\end{equation}
By substituting then $\mathrm{e}^{\chi / \rho^2} = x$, one obtains
\begin{equation}
\Xi (\eta_-, \eta_+, \varepsilon, \beta) = \pi \beta \int \limits_1^{+\infty} \frac{dx}{x^3} \, \mathrm{exp} \left \{ - \frac{1}{\varepsilon \sqrt{f(\eta_-) f(\eta_+)}} \big [ \rho (\eta_+ - \eta_-, \beta) x - 1 \big ] \right \}.
\label{eq:app_Xi_x}
\end{equation}
Note that if we replace the function $\Xi$ in Eq.~\eqref{eq:app_nu_G} with unity, the expression will yield $\nu^{(\text{PWA})}$ [see Eq.~\eqref{eq:pwa_final}]. To identify the transverse volume factor, we first calculate the integral in Eq.~\eqref{eq:app_Xi_x}:
\begin{equation}
\Xi (\eta_-, \eta_+, \varepsilon, \beta) = \pi \beta \, \mathrm{e}^{1/q} \int \limits_1^{+\infty} \frac{dx}{x^3} \, \mathrm{e}^{-x/\mu} = \frac{\pi \beta}{2} \, \mathrm{e}^{1/q - 1/\mu} \bigg [ 1 - \frac{1}{\mu} - \frac{\mathrm{e}^{1/\mu}}{\mu^2} \, \mathrm{Ei} \, (-1/\mu) \bigg ],
\label{eq:app_Xi_integral}
\end{equation}
where $q \equiv \varepsilon \sqrt{f(\eta_-) f(\eta_+)}$ and $\mu \equiv q/\rho (\eta_+ - \eta_-, \beta)$. Since $\eta_\pm$ are always small or of the order of unity (if we consider the side maxima of the external field) and $\beta \gg 1$, the function $\rho (\eta_+ - \eta_-, \beta)$ is always very close to $1$. Accordingly, $q$ and $\mu$ are of the same order and small in the weak-field limit. Let us expand Eq.~\eqref{eq:app_Xi_integral} as follows:
\begin{equation}
\Xi (\eta_-, \eta_+, \varepsilon, \beta) = \pi \beta \, \mathrm{e}^{1/q - 1/\mu} \mu \big [ 1 - 3\mu + \mathcal{O} (\mu^2) \big ].
\label{eq:app_Xi_expansion}
\end{equation}
Although it is already clear that the simplest approximation for $\nu^{(\text{G})}/\nu^{(\text{PWA})}$ has the form $\pi \beta \varepsilon = (\omega w_0)^2 (E_0/E_\text{c})$, we will perform more accurate calculations. To this end, we first expand $\mu$ similarly to Eq.~\eqref{eq:series_f}:
\begin{equation}
\mu = \varepsilon \bigg [ 1 - \frac{2+\sigma^2}{4\sigma^2} \, \big (\eta_-^2 + \eta_+^2 \big )  + \mathcal{O} (\eta^4) \bigg ].
\end{equation}
We also need
\begin{equation}
\frac{1}{\mu} = \frac{1}{\varepsilon} \, \bigg [ 1 + \frac{2+\sigma^2}{4\sigma^2} \, \big (\eta_-^2 + \eta_+^2 \big ) + \frac{12+12\sigma^2+7\sigma^4}{96\sigma^4} \, \big (\eta_-^4 + \eta_+^4 \big ) + \frac{(2+\sigma^2)^2}{16\sigma^4} \, \eta_-^2 \eta_+^2 + \mathcal{O} (\eta^6) \bigg ].
\end{equation}
The first term gives rise to the main exponential factor $\mathrm{e}^{-\pi E_\text{c}/E_0}$. The second term determines a Gaussian profile to be integrated over $\eta_\pm$ in Eq.~\eqref{eq:app_nu_G}. Finally, the two subsequent terms of the order of $\eta^4$ affect the $\mathcal{O}(E_0/E_\text{c})$ contribution to the prefactor in our estimate. By performing the integration and collecting the terms, we find
\begin{equation}
\nu^{(\text{G})} \approx \frac{1}{2\pi^3} \, (\omega w_0)^2 \, \frac{\sigma^2}{2+\sigma^2} \bigg(\frac{E_0}{E_ \text{c}}\bigg)^4 \mathrm{e}^{- \pi E_\text{c}/E_0} \bigg [1-\frac{4(7+7\sigma^2+2\sigma^4)}{\pi (2+\sigma^2)^2} \frac{E_0}{E_\text{c}}+\mathcal{O} \bigg(\frac{E_0^2}{E_\text{c}^2}\bigg)\bigg].
\end{equation}
We will again approximate the linear-term coefficient in square brackets as $-8/\pi + \mathcal{O} (1/\sigma^2)$. By adding now the side-maxima contributions from $\eta = \pm \pi$, we obtain Eq.~\eqref{eq:G_laplace}.

\end{widetext}



\begin{thebibliography}{99}
%
\bibitem{sauter_1931} F.~Sauter, \"Uber das Verhalten eines Elektrons im homogenen elektrischen Feld nach der relativistischen Theorie Diracs, Z.~Phys. {\bf 69}, 742 (1931).
%
\bibitem{euler_kockel} H.~Euler and B.~Kockel, \"Uber die Streuung von Licht an Licht nach der Diracschen Theorie, Naturwiss. {\bf 23}, 246 (1935).
%
\bibitem{heisenberg_euler} W.~Heisenberg and H.~Euler, Folgerungen aus der Diracschen Theorie des Positrons, Z.~Phys. {\bf 98}, 714 (1936).
%
\bibitem{schwinger_1951} J.~Schwinger, On gauge invariance and vacuum polarization, Phys. Rev. {\bf 82}, 664 (1951).
%
\bibitem{gonoskov_2022} A.~Gonoskov, T.~G.~Blackburn, M.~Marklund, and S.~S.~Bulanov, Charged particle motion and radiation in strong electromagnetic fields, Rev. Mod. Phys. {\bf 94}, 045001 (2022).
%
\bibitem{fedotov_review} A.~Fedotov, A.~Ilderton, F.~Karbstein, B.~King, D.~Seipt, H.~Taya, and G.~Torgrimsson, Advances in QED with intense background fields, Phys. Rep. {\bf 1010}, 1 (2023).
%
\bibitem{fradkin_gitman_shvartsman} E.~S.~Fradkin, D.~M.~Gitman, and S.~M.~Shvartsman, {\it Quantum Electrodynamics with Unstable Vacuum} (Springer-Verlag, Berlin, 1991).
%
\bibitem{BB_prd_1991} I.~Bialynicki-Birula, P.~G\'ornicki, and J.~Rafelski, Phase-space structure of the Dirac vacuum, Phys. Rev. D {\bf 44}, 1825 (1991).
%
\bibitem{gavrilov_prd_1996} S.~P.~Gavrilov and D.~M.~Gitman, Vacuum instability in external fields, Phys. Rev. D {\bf 53}, 7162 (1996).
%
\bibitem{zhuang_1996} P.~Zhuang and U.~Heinz, Relativistic quantum transport theory for electrodynamics, Ann. Phys. {\bf 245}, 311 (1996).
%
\bibitem{ochs_1998} S.~Ochs and U.~Heinz, Wigner functions in covariant and single-time formulations, Ann. Phys. {\bf 266}, 351 (1998).
%
\bibitem{schmidt_1998} S.~Schmidt, D.~Blaschke, G.~R\"opke, S.~A.~Smolyansky, A.~V.~Prozorkevich, and V.~D.~Toneev, A quantum kinetic equation for particle production in the Schwinger mechanism, Int. J. Mod. Phys. E {\bf 07}, 709 (1998).
%
\bibitem{kluger_prd_1998} Y.~Kluger, E.~Mottola, and J.~M.~Eisenberg, Quantum Vlasov equation and its Markov limit, Phys. Rev. D {\bf 58}, 125015 (1998).
%
\bibitem{pervushin_skokov} V.~N.~Pervushin and V.~V.~Skokov, Kinetic description of fermion production in the oscillator representation, Acta Phys. Polon. B {\bf 37}, 2587 (2006).
%
\bibitem{hebenstreit_prd_2010} F.~Hebenstreit, R.~Alkofer, and H.~Gies, Schwinger pair production in space- and time-dependent electric fields: Relating the Wigner formalism to quantum kinetic theory, Phys. Rev. D {\bf 82}, 105026 (2010).
%
\bibitem{blaschke_prd_2011} D.~B.~Blaschke, V.~V.~Dmitriev, G.~R\"opke, and S.~A.~Smolyansky, BBGKY kinetic approach for an $e^-e^+\gamma$ plasma created from the vacuum in a strong laser-generated electric field: The one-photon annihilation channel, Phys. Rev. D {\bf 84}, 085028 (2011).
%
\bibitem{blinne_gies_2014} A.~Blinne and H.~Gies, Pair production in rotating electric fields, Phys. Rev. D {\bf 89}, 085001 (2014).
%
\bibitem{woellert_prd_2015} A.~W\"ollert, H.~Bauke, and C.~H.~Keitel, Spin polarized electron-positron pair production via elliptical polarized laser fields, Phys. Rev. D {\bf 91}, 125026 (2015).
%
\bibitem{blinne_strobel_2016} A.~Blinne and E.~Strobel, Evolution of chirality in a multiphoton pair production process, Phys. Rev. D {\bf 93}, 025014 (2016).
%
\bibitem{aleksandrov_prd_2016} I.~A.~Aleksandrov, G.~Plunien, and V.~M.~Shabaev, Electron-positron pair production in external electric fields varying both in space and time, Phys. Rev. D {\bf 94}, 065024 (2016).
%
\bibitem{li_prd_2017} Z.~L.~Li, Y.~J.~Li, and B.~S.~Xie, Momentum vortices on pairs production by two counter-rotating fields, Phys. Rev. D {\bf 96}, 076010 (2017).
%
\bibitem{lv_pra_2018} Q.~Z.~Lv, S.~Dong, Y.~T.~Li, Z.~M.~Sheng, Q.~Su, and R.~Grobe, Role of the spatial inhomogeneity on the laser-induced vacuum decay, Phys. Rev. A {\bf 97}, 022515 (2018).
%
\bibitem{schneider_prd_2018} C.~Schneider, G.~Torgrimsson, and R.~Sch\"utzhold, Discrete worldline instantons, Phys. Rev. D {\bf 98}, 085009 (2018).
%
\bibitem{huang_2019} X.~G.~Huang, M.~Matsuo, and H.~Taya, Spontaneous generation of spin current from the vacuum by strong electric fields, Prog. Theor. Exp. Phys. {\bf 2019}, 113B02 (2019)
%
\bibitem{huang_prd_2019} X.~G.~Huang and H.~Taya, Spin-dependent dynamically assisted Schwinger mechanism, Phys. Rev. D {\bf 100}, 016013 (2019).
%
\bibitem{aleksandrov_epjst_2020} I.~A.~Aleksandrov, V.~V.~Dmitriev, D.~G.~Sevostyanov, and S.~A.~Smolyansky, Kinetic description of vacuum $e^+e^-$ production in strong electric fields of arbitrary polarization, Eur. Phys. J. Spec. Top. {\bf 229}, 3469 (2020).
%
\bibitem{aleksandrov_kohlfuerst} I.~A.~Aleksandrov and C.~Kohlf\"urst, Pair production in temporally and spatially oscillating fields, Phys. Rev. D {\bf 101}, 096009 (2020).
%
\bibitem{aleksandrov_kudlis_klochai} I.~A.~Aleksandrov, A.~Kudlis, and A.~I.~Klochai, Kinetic theory of vacuum pair production in uniform electric fields revisited, Phys. Rev. Res. {\bf 6}, 043009 (2024).
%
\bibitem{esposti_prd_2024} G.~Degli Esposti and G.~Torgrimsson, Momentum spectrum of Schwinger pair production in four-dimensional e-dipole fields, Phys. Rev. D {\bf 109}, 016013 (2024).
%
\bibitem{ruf_prl_2009} M.~Ruf, G.~R.~Mocken, C.~M\"uller, K.~Z.~Hatsagortsyan, and C.~H.~Keitel, Pair production in laser fields oscillating in space and time, Phys. Rev. Lett. {\bf 102}, 080402 (2009).
%
\bibitem{aleksandrov_2017_2} I.~A.~Aleksandrov, G.~Plunien, and V.~M.~Shabaev, Momentum distribution of particles created in space-time-dependent colliding laser pulses, Phys. Rev. D {\bf 96}, 076006 (2017).
%
\bibitem{kohlfuerst_prl_2022} C.~Kohlf\"urst, N.~Ahmadiniaz, J.~Oertel, and R.~Sch\"utzhold, Sauter-Schwinger effect for colliding laser pulses, Phys. Rev. Lett. {\bf 129}, 241801 (2022).
%
\bibitem{kohlfuerst_arxiv_2022_2} C.~Kohlf\"urst, Pair production in circularly polarized waves, Phys. Rev. D {\bf 110}, L111903 (2024).
%
\bibitem{aleksandrov_prd_2017_1} I.~A.~Aleksandrov, G.~Plunien, and V.~M.~Shabaev, Pulse shape effects on the electron-positron pair production in strong laser fields, Phys. Rev. D {\bf 95}, 056013 (2017).
%
\bibitem{olugh_prd_2019} O.~Olugh, Z.~L.~Li, B.~S.~Xie, and R.~Alkofer, Pair production in differently polarized electric fields with frequency chirps, Phys. Rev. D {\bf 99}, 036003 (2019).
%
\bibitem{kohlfuerst_prd_2019} C.~Kohlf\"urst, Spin states in multiphoton pair production for circularly polarized light, Phys. Rev. D {\bf 99}, 096017 (2019).
%
\bibitem{hu_arxiv_2024} L.~N.~Hu, H.~H.~Fan, O.~Amat, S.~Tang, and B.~S.~Xie, Spin effect induced momentum spiral and asymmetry degree in pair production, Phys. Rev. D {\bf 110}, 056013 (2024).
%
\bibitem{majczak_arxiv_2024} M.~M.~Majczak, K.~Krajewska, J.~Z.~Kami\ifmmode \acute{n}\else \'{n}\fi{}ski, and A.~Bechler, Scattering matrix approach to dynamical Sauter-Schwinger process: Spin- and helicity-resolved momentum distributions, Phys. Rev. D {\bf 110}, 116025 (2024).
%
\bibitem{aleksandrov_kudlis} I.~A.~Aleksandrov and A.~Kudlis, Pair production in rotating electric fields via quantum kinetic equations: Resolving helicity states, Phys. Rev. D {\bf 110}, L011901 (2024).
%
\bibitem{bunkin_tugov} F.~V.~Bunkin and I.~I.~Tugov, Possibility of creating electron-positron pairs in a vacuum by the focusing of laser radiation, Dokl. Akad. Nauk SSSR {\bf 187}, 541 (1969) [Sov. Phys. Dokl. {\bf 14}, 678 (1970)].
%
\bibitem{narozhny_pla_2004} N.~B.~Narozhny, S.~S.~Bulanov, V.~D.~Mur, and V.~S.~Popov, $e^+e^-$-pair production by a focused laser pulse in vacuum, Phys. Lett. A {\bf 330}, 1 (2004).
%
\bibitem{narozhny_jetpl_2004} N.~B.~Narozhny, S.~S.~Bulanov, V.~D.~Mur, and V.~S.~Popov, On $e^+e^-$ pair production by colliding electromagnetic pulses, JETP Lett. {\bf 80}, 382 (2004).
%
\bibitem{bulanov_jetp_2006} S.~S.~Bulanov, N.~B.~Narozhny, V.~D.~Mur, and V.~S.~Popov, Electron-positron pair production by electromagnetic pulses, JETP {\bf 102}, 9 (2006).
%
\bibitem{dunne_prd_2006} G.~V.~Dunne, Q.~H.~Wang, H.~Gies, and C.~Schubert, Worldline instantons and the fluctuation prefactor, Phys. Rev. D {\bf 73}, 065028 (2006).
%
\bibitem{hebenstreit_prdr_2008} F.~Hebenstreit, R.~Alkofer, and H.~Gies, Pair production beyond the Schwinger formula in time-dependent electric fields, Phys. Rev. D {\bf 78}, 061701(R) (2008).
%
\bibitem{bulanov_prl_2010} S.~S.~Bulanov, V.~D.~Mur, N.~B.~Narozhny, J.~Nees, and V.~S.~Popov, Multiple colliding electromagnetic pulses: A way to lower the threshold of ${e}^{+}{e}^{\ensuremath{-}}$ pair production from vacuum, Phys. Rev. Lett. {\bf 104}, 220404 (2010).
%
\bibitem{gavrilov_prd_2017} S.~P.~Gavrilov and D.~M.~Gitman, Vacuum instability in slowly varying electric fields, Phys. Rev. D {\bf 95}, 076013 (2017).
%
\bibitem{aleksandrov_prd_2019_1} I.~A.~Aleksandrov, G.~Plunien, and V.~M.~Shabaev, Locally-constant field approximation in studies of electron-positron pair production in strong external fields, Phys. Rev. D {\bf 99}, 016020 (2019).
%
\bibitem{sevostyanov_prd_2021} D.~G.~Sevostyanov, I.~A.~Aleksandrov, G.~Plunien, and V.~M.~Shabaev, Total yield of electron-positron pairs produced from vacuum in strong electromagnetic fields: Validity of the locally constant field approximation, Phys. Rev. D {\bf 104}, 076014 (2021).
%
\bibitem{aleksandrov_symmetry_2022} I.~A.~Aleksandrov, D.~G.~Sevostyanov, and V.~M.~Shabaev, Particle production in strong electromagnetic fields and local approximations, Symmetry {\bf 14}, 2444 (2022).
%
\bibitem{aleksandrov_prd_2018} I.~A. Aleksandrov, G.~Plunien, and V.~M. Shabaev, Dynamically assisted Schwinger effect beyond the spatially-uniform-field approximation, Phys. Rev. D. {\bf 97}, 116001 (2018).
%
\bibitem{peng_prr_2020} Z.~Peng, H.~Hu, and J.~Yuan, Multichannel interference in nonperturbative multiphoton pair production by gamma rays colliding, Phys. Rev. Res. {\bf 2}, 013020 (2020).
%
\bibitem{nikishov_constant} A.~I.~Nikishov, Pair production by a constant external field, Zh. Eksp. Teor. Fiz. {\bf 57}, 1210 (1969) [Sov. Phys. JETP {\bf 30}, 660 (1970)].
%
\bibitem{schuetzhold_prl_2008} R.~Sch\"utzhold, H.~Gies, and G.~Dunne, Dynamically assisted Schwinger mechanism, Phys. Rev. Lett. {\bf 101}, 130404 (2008).
%
\bibitem{ilderton_prd_2022} A.~Ilderton, Physics of adiabatic particle number in the Schwinger effect, Phys. Rev. D {\bf 105}, 016021 (2022).
%
\bibitem{aleksandrov_dase_2022} I.~A.~Aleksandrov, D.~G.~Sevostyanov, and V.~M.~Shabaev, Schwinger particle production: Rapid switch off of the external field versus dynamical assistance, Phys. Rev. D {\bf 111}, 016010 (2025).
%
\bibitem{gavrilov_prd_2016_gen} S.~P.~Gavrilov and D.~M.~Gitman, Quantization of charged fields in the presence of critical potential steps, Phys. Rev. D {\bf 93}, 045002 (2016).
%
\end{thebibliography}
\end{document}